\documentclass{article}

\usepackage{arxiv}

\usepackage{newtxtext,newtxmath}
\usepackage{graphicx}
\usepackage{amsmath}
\usepackage[utf8]{inputenc} 
\usepackage[T1]{fontenc}    
\usepackage{hyperref}       
\usepackage{url}            
\usepackage{booktabs}       
\usepackage{amsfonts}       
\usepackage{nicefrac}       
\usepackage{microtype}      
\usepackage{lipsum}		
\usepackage{graphicx}
\usepackage[numbers,sort&compress]{natbib}

\usepackage{doi}
\usepackage{float}
\usepackage{afterpage}

\makeatletter
\renewcommand{\fnum@figure}{\textbf{Fig. \thefigure}}
\renewcommand{\fnum@table}{\textbf{Table \thetable}}
\makeatother

\usepackage[labelformat=empty]{caption}
\usepackage[displaymath,mathlines]{lineno}
\usepackage{siunitx}
\usepackage{wasysym}
\usepackage{titling}
\usepackage[normalem]{ulem}
\usepackage{xcolor}
\usepackage{titlesec}
\usepackage{url}
\hypersetup{
    hidelinks
}

\definecolor{headingcolor}{HTML}{792F3A}

\titleformat{\section}
  {\Large\bfseries\raggedright\color{headingcolor}}
  {\thesection}{1em}{}

\titleformat{\subsection}
  {\large\bfseries\raggedright\color{headingcolor}}
  {\thesubsection}{1em}{}

\titleformat{\subsubsection}
  {\normalsize\itshape\color{headingcolor}}
  {\thesubsubsection}{1em}{}

\usepackage{titlesec}

\titlespacing*{\section}{0pt}{12pt}{-4pt}
\titlespacing*{\subsection}{0pt}{10pt}{-4pt}
\titlespacing*{\subsubsection}{0pt}{8pt}{-4pt}

\date{}

\def\scititle{Metamaterial tactile displays with \\programmable virtual pixels}

\title{\bfseries\boldmath\scititle}

\author{
    Gregory~Reardon$^{1}$, Max Linnander$^{2}$, Dustin Goetz$^{2}$,\\ Neeli Tummala$^{3}$, and Yon Visell$^{2\ast}$\\
    \small{$^1$Media Arts and Technology Program, $^2$Department of Mechanical Engineering,}\\
    \small{$^3$Department of Electrical and Computer Engineering,} \\
    \small{University of California, Santa Barbara, USA.}\\
    \small{$^\ast$Corresponding author. E-mail:  yonvisell@ucsb.edu.}
}

\begin{document} 

\captionsetup[figure]{font={scriptsize,sf}}
\sisetup{inter-unit-product = \ensuremath{{\cdot}}} 

\maketitle 

\begin{abstract} \bfseries \boldmath

Tactile displays present information by supplying localized mechanical signals across their surface, but conventional approaches often require an actuator to be positioned at every location where feedback is provided, limiting resolution and increasing complexity. Wave-mediated tactile displays exploit wave transmission to generate virtual pixels using far fewer actuators, but are constrained in resolution by the intrinsic dispersion characteristics of the display medium. Here, we show how this constraint can be overcome by applying metamaterial physics to engineer wave transmission in a rigid plate using an optimized lattice of resonators. The resulting system exhibits a slow-wave dispersion branch that reduces the wavelengths available for constructing virtual pixels, thereby enhancing display resolution by an order of magnitude. We demonstrate a metamaterial tactile display driven by eight peripheral actuators that is capable of generating localized virtual tactile pixels at any surface location with centimeter-scale resolution. The location, timing, and waveform of each pixel can be specified computationally, and the output bandwidth, 75--400\,Hz, spans a substantial portion of the tactile frequency range. In four perceptual experiments, participants localized, explored, and discriminated multi-pixel patterns comprising transient and sustained vibrations with accuracies ranging from 95.3\% to 100\%. This research provides metamaterial engineering methods that enable the design of tactile displays driven by few actuated degrees of freedom, with implications for many future application areas.

\end{abstract}

\section*{One-Sentence Summary}
Metamaterial tactile displays present dynamic, localized virtual pixels using sparse actuator arrays.

\section*{INTRODUCTION}

A central goal of tactile display design is to present digital information through touch, allowing users to feel and interact with otherwise intangible content. To support this goal, tactile displays should be capable of generating programmable mechanical signals at multitudes of locations across the display. These capabilities would allow tactile feedback to be spatially organized and temporally sequenced, enabling digital information to be conveyed as distributed and dynamic tactile patterns. Potential applications span human-machine interfaces~\cite{hayward2004haptic,basdogan2020review}, virtual and augmented reality~\cite{culbertson2018haptics, wang2019multimodal}, data visualization~\cite{paneels2009review}, and robotic teleoperation~\cite{pacchierotti2023cutaneous}.

Despite sustained research efforts, generating distributed tactile feedback remains challenging, as displays must simultaneously provide fine spatial localization, wide temporal bandwidth, and sufficient dynamic range while maintaining manageable system complexity. Conventional approaches typically employ architectures in which one actuator is used to drive each location on the display surface~\cite{ikei1997vibratory, ujitoko2020development, shen2023fluid, purnendu2023fingertip, youn2025skin, tan2025toward, linnander2026thermopneumatic}. Even with advances in functional materials and actuation technologies, fabricating high-resolution displays using dense actuator arrays introduces substantial challenges in integration, wiring, packaging, power delivery, and control~\cite{biswas2019emerging}. Many displays that leverage one-actuator-to-one-pixel architectures deliver tactile feedback over small interaction areas, as in wearable fingertip displays; this approach reduces demands on actuator count. However, expanding the pixel count to support surface-based tactile displays, where users can interact across larger areas without worn hardware, exacerbates the burdens associated with actuator integration density, negatively impacting practicality and manufacturability.

\begin{figure*}[t!]
    \hspace*{-2mm}
    \includegraphics[scale=0.9]{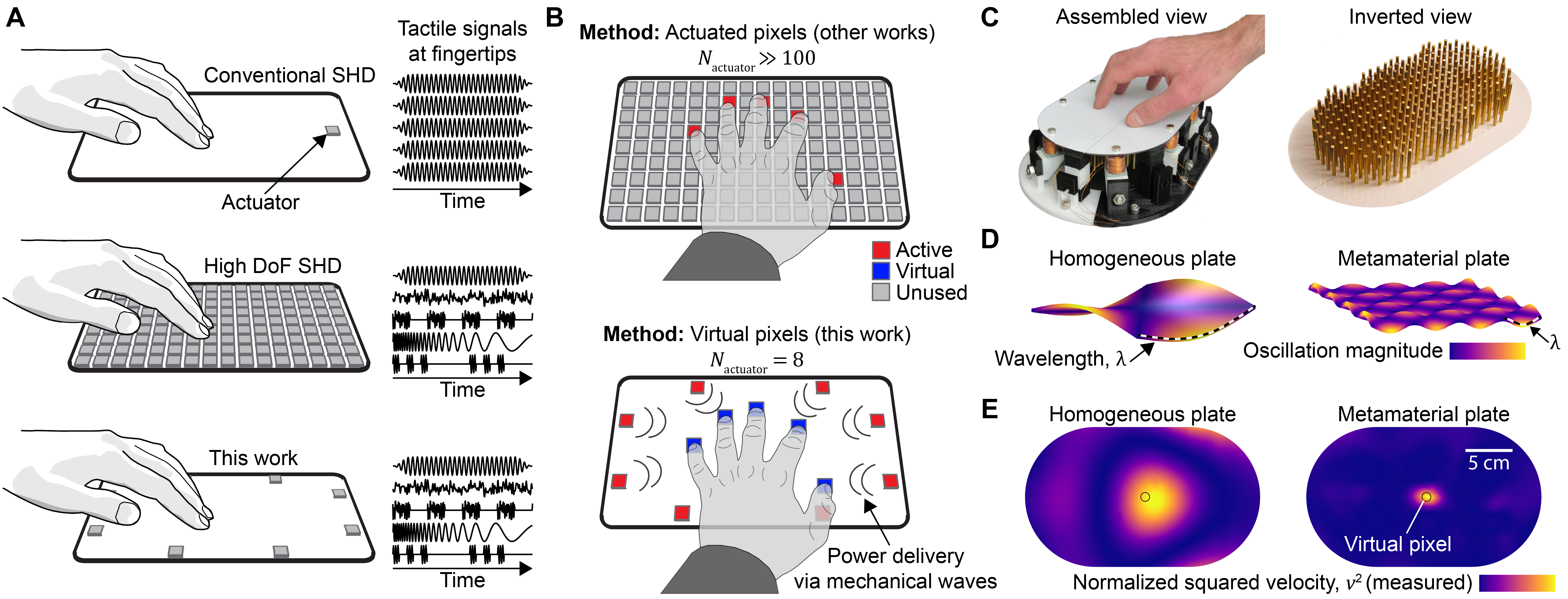}
    \caption{\textbf{Fig. 1. Computational wave-field control using locally resonant metamaterials for tactile displays.} (\textbf{A}) Conventional tactile displays supply uniform tactile feedback across the display with a single actuated degree of freedom (DOF), yielding similar temporal waveforms at the fingertip contact locations (top; signal traces for each fingertip shown at right). Independent control over the temporal waveform delivered to each finger can be achieved either with high-DOF devices (middle) or sparse actuation combined with computational wave-field control, as used here (bottom). (\textbf{B}) High-DOF devices drive localized surface regions with independent actuators (top). Our approach instead forms virtual pixels by remotely focusing mechanical waves, enabling dynamic control and reconfiguration of the pixels (bottom). (C) Assembled metamaterial tactile display (left), driven by eight peripherally located electromagnetic actuators. The metamaterial plate consists of a thin acrylic plate ($h = 1.5$ mm) coupled to a square lattice ($a = 8$ mm) of brass rod resonators ($h_{\mathrm{rod}} = 33.46$ mm, $d_{\mathrm{rod}} = 3.175$ mm). The plate is shown inverted (right) to expose the resonators. (\textbf{D}) Visualizations of waves on a homogeneous plate (left) and a metamaterial plate (right), with wavelengths $\lambda$. The smaller wavelengths on the metamaterial plate facilitate the generation of highly localized virtual pixels. (\textbf{E}) Experimental results (shown here and detailed in later sections) demonstrate that waves can be controlled to generate virtual pixels with areas as small as \SI{2}{\centi\meter\squared} on the metamaterial plate (right), an order of magnitude smaller than those generated on an otherwise identical plate without resonators (left).}
    \label{fig:metaplate-concept}
\end{figure*}

An emerging alternative is to leverage a comparatively small number of actuators to focus mechanical waves within a display at multiple locations (Fig.~\ref{fig:metaplate-concept}A). These focal regions can be interpreted as virtual pixels (Fig.~\ref{fig:metaplate-concept}B), whose spatial and temporal characteristics are computationally programmable and dynamically reconfigurable to match the requirements of hand-surface interactions~\cite{bai2011impact, hudin2015localized, woo2015vibration, emgin2018haptable, hudin2018localisation, pantera2020multitouch, reardon2020elastowave, goetz2023dynamic, reardon2023rendering, hudin2026spatio}. Compared with displays using dense actuator arrays, these wave-based methods can generate spatially localized tactile feedback while typically requiring one to two orders of magnitude fewer independently controlled actuator inputs (actuated degrees of freedom, or DOFs), substantially reducing system complexity. A range of computational wave-field control methods have been investigated for shaping the spatiotemporal evolution of plate waves to generate localized vibrotactile or friction-modulation feedback, including modal superposition~\cite{woo2015vibration, katumu2016using}, time-reversal focusing~\cite{hudin2015localized, reardon2020elastowave}, and spatiotemporal inverse filtering~\cite{pantera2020multitouch}. These methods build on foundational techniques in several related fields, such as multi-speaker audio systems for spatial and immersive audio~\cite{gerzon1985ambisonics,berkhout1993acoustic, pulkki1997virtual}, mid-air holographic haptic displays~\cite{hoshi2010noncontact,carter2013ultrahaptics}, and ultrasound-based medical imaging and treatment~\cite{fink1992time, tanter2001optimal, aubry2001optimal, pernot2007vivo, ter2007high, gennisson2013ultrasound, taljanovic2017shear, izadifar2020introduction}.

Although existing methods provide powerful tools for shaping plate waves, localized actuation on continuum surfaces remains fundamentally limited by the underlying wave physics. The dispersion and loss characteristics of the display surface constrain the achievable localization and, thus, display resolution, irrespective of the computational focusing technique. The dispersion relation determines the frequency-dependent phase velocity, $c(\omega)$, and therefore the spatial wavelength of mechanical energy, $\lambda(\omega)=2\pi c(\omega)/\omega$. Consequently, focusing resolution is bounded by the diffraction limit, $d_\mathrm{res} (\omega) \gtrsim \lambda(\omega) / 2$, preventing the generation of virtual tactile pixels below this scale. At tactile frequencies ($\approx$ \num{20}--\SI{800}{\hertz}~\cite{sato1961response,johansson1982responses}), stiff plates support flexural waves that propagate with high phase velocities and correspondingly long wavelengths, yielding virtual pixels with diameters of \SI{10}{\centi\meter} or more. In contrast, soft elastomers support shear waves that propagate at lower phase velocities, enabling finer resolution, with virtual pixel diameters on the order of \SI{1}{\centi\meter}~\cite{goetz2023dynamic,reardon2023rendering}. However, soft elastomers are often less robust to repeated touch interactions and exhibit high viscoelastic losses at tactile frequencies, degrading the fidelity of virtual pixels through non-recoverable energy dissipation and thereby requiring increased actuation density and closer placement of actuators relative to the virtual pixels.

Here, we report a metamaterial-based approach to overcoming these limitations and enabling high-resolution, low-DOF wave-based tactile displays on stiff plates. We achieve this by coupling the plate to a periodic array of local resonators (Fig.~\ref{fig:metaplate-concept}C), altering the dispersion relation governing wave propagation. As a result, our metamaterial plate enables spatial control of wave fields beyond the diffraction limit imposed by a homogeneous plate~\cite{lemoult2011acoustic, rupin2014experimental}, supporting lower phase velocities and shorter wavelengths at tactile frequencies (Fig.~\ref{fig:metaplate-concept}D). Our approach leverages established principles from locally resonant acoustic and elastic metamaterials, which have been explored for low-frequency sound and vibration attenuation~\cite{liu2000locally, yang2008membrane, mei2012dark, peng2015acoustic}, energy localization and harvesting~\cite{chen2014metamaterials, li2017design, lee2022piezoelectric}, and resonant subwavelength imaging~\cite{li2009experimental, zhu2011holey, kaina2015negative}. We further combine dispersion engineering with active wave-field control, using the metamaterial not only to passively modify wave propagation but also as a programmable medium for shaping the spatiotemporal evolution of plate waves. This substantially enhances the resolution of tactile feedback on a stiff substrate (Fig.~\ref{fig:metaplate-concept}E).

We demonstrate a metamaterial tactile display that can generate multiple dynamically reconfigurable virtual pixels across a large touch surface ($>$ \SI{300}{\centi\meter\squared}) with centimeter-scale focal resolution. We achieve this using a sparse, peripheral array of eight actuators to computationally control surface waves on the metamaterial plate via spatiotemporal inverse filtering. This approach enables the location, timing, and temporal waveform of each virtual pixel to be specified independently, supporting both transient and sustained vibrations. The resulting virtual pixels are spatially localized, operate over a broad portion of the tactile frequency range (\num{75}--\SI{400}{\hertz}), and deliver peak velocities exceeding \SI{10}{\milli\meter/\second} (with peak displacements of approximately \SI{10}{\micro\metre}), more than two orders of magnitude above perceptual thresholds~\cite{verrillo1963effect,verrillo1971vibrotactile}. Across four perceptual experiments, participants were able to localize, explore, and discriminate single- and multi-pixel patterns with accuracies ranging from 95.3\% to 100\%. These results demonstrate that dispersion engineering using metamaterials can support spatially localized, broadband, and independently addressable tactile feedback using few actuated DOFs.

\section*{RESULTS}

To overcome material and diffraction limits on wave focusing, we realize a metamaterial display by coupling a continuum surface to an ensemble of identical mechanical resonators. In a homogeneous (unmodified) plate, waves propagate at phase velocities that are too high to produce virtual pixels at spatial scales comparable to those of a fingerpad. Even in thin plates ($\approx \SI{1.5}{\milli\meter}$) with relatively low stiffness (Young's modulus of \SI{3}{\giga\pascal}; plate flexural rigidity of \SI{1}{\newton\meter}), these waves ($\mathrm{A}_0$ mode) have wavelengths on the order of tens of centimeters or more. Coupling the resonators to the plate introduces additional mass and modifies the plate’s dispersion relation through mode hybridization. Together, these effects shift the relevant characteristic frequencies downward, thereby reducing phase velocity and wavelength within the tactile frequency range. As a result, wave fields can be focused into highly localized
surface regions that serve as virtual tactile pixels.

To connect mode hybridization to wavelength reduction, we analyze the dispersion of the periodic plate--resonator system along a given propagation direction with scalar wavenumber $k=|\mathbf{k}|$. At each fixed $k$, an uncoupled plate branch with frequency $\omega_0$ couples to a resonator-associated mode with frequency $\omega_r$, yielding the characteristic equation
\begin{equation}
\det
\begin{bmatrix}
\omega_0^2 - \omega^2 & \varepsilon \\
\varepsilon & \omega_r^2 - \omega^2
\end{bmatrix}
= 0,
\end{equation}
where $\varepsilon$ characterizes the coupling strength at the given wavenumber $k$. The characteristic equation is solved for the two allowed squared frequencies, $\omega^2$. Writing these roots relative to the uncoupled plate squared frequency, i.e., $\omega^2 = \omega_0^2 + \Delta$, yields
\begin{equation}
\Delta_{\pm} = 
\frac{1}{2}
\left(
\omega_r^2 - \omega_0^2
\pm
\sqrt{\left[\omega_r^2 - \omega_0^2\right]^2 + 4 \varepsilon^2}
\right),
\end{equation}
with corresponding eigenvalues $\omega_{\pm}^2 = \omega_0^2 + \Delta_{\pm}$. Repeating this calculation for each $k$ yields the hybridized dispersion branches $\omega_{\pm}^2(k) = \omega_0^2(k) + \Delta_{\pm}(k)$. The lower solution, $\omega_{-}^2(k)$, is shifted downward relative to the uncoupled plate branch, forming a slow-wave branch of the dispersion relation. At a fixed operating frequency $\omega$ on this branch, the corresponding wavenumber $k(\omega)$ is larger than in the uncoupled plate, producing a lower phase velocity $c(\omega)=\omega/k(\omega)$ and a shorter wavelength $\lambda(\omega)=2\pi/k(\omega)$. Because focusing resolution is constrained by the diffraction limit, $d_\mathrm{res}(\omega) \gtrsim \lambda(\omega)/2$, this wavelength reduction improves the achievable display resolution. In a finite plate, this same dispersion shift can increase the number of distinct spatial response patterns within the operating bandwidth, thereby enhancing wave-field control~\cite{etaix2013increasing,dupre2015wave}.

\begin{figure*}[t!]
    \centering
    \includegraphics[scale=0.9]{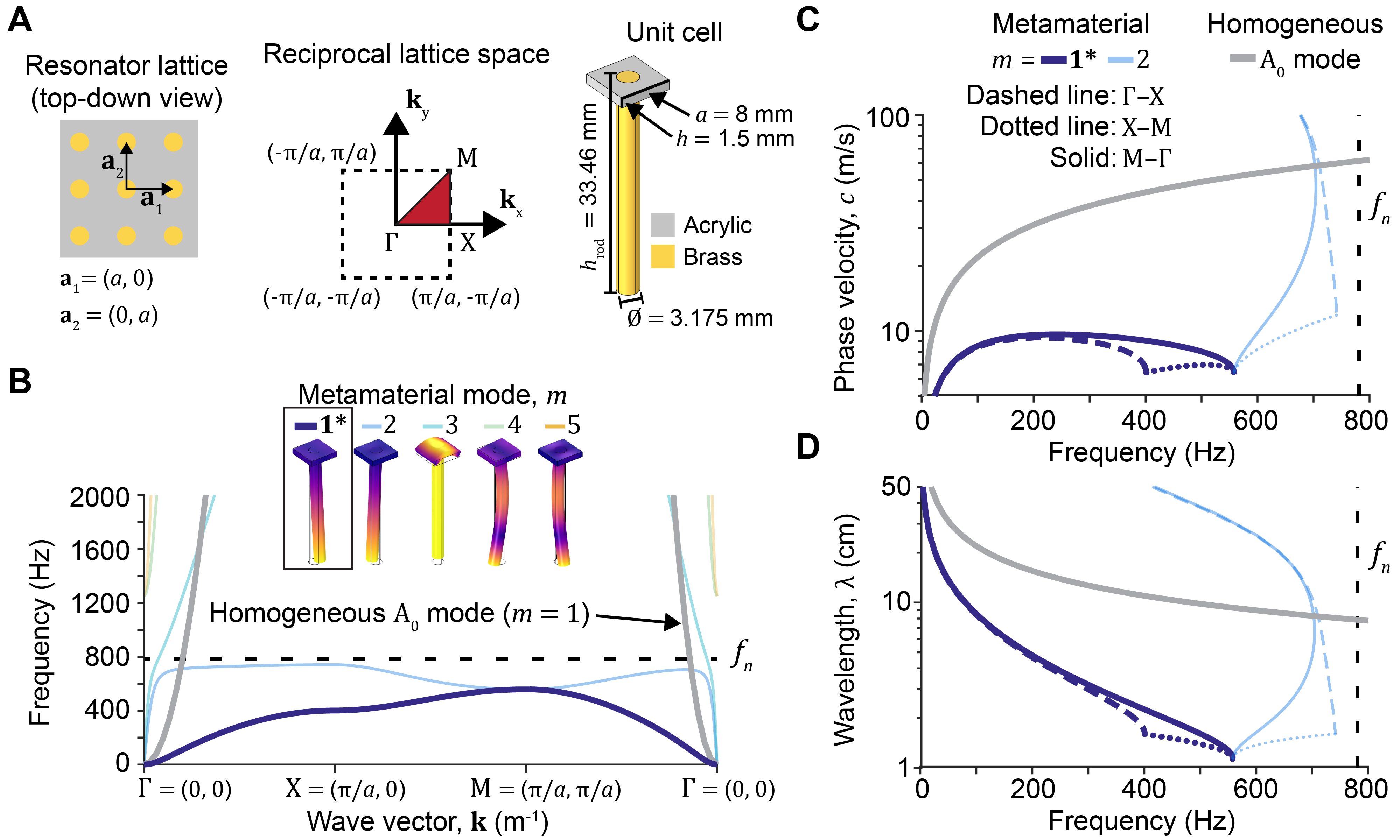}
    \caption{\textbf{Fig. 2. Surface mode hybridization yields a slow-wave branch in a numerically simulated locally resonant metamaterial.} (\textbf{A}) Dispersion was computed using numerical simulations of an infinite periodic square lattice (left), evaluated along a reciprocal-lattice path defined by the high-symmetry points $\Gamma$, $\mathrm{X}$, and $\mathrm{M}$ (middle). Design parameters (right) were iteratively optimized to produce a slow-wave branch overlapping most of the tactile frequency range. (\textbf{B}) Coupling between plate and resonator modes yielded a slow-wave branch associated with the first eigenmode solution ($m=1$). First five modes shown; inset: mode shapes evaluated at $\mathbf{k}=\mathrm{X}$; gray dashed line: $\mathrm{A}_0$ mode ($m=1$) of the homogeneous plate; black line: frequency $f_n$ of the first resonator-associated mode. (\textbf{C}) Phase velocity $c$ and (\textbf{D}) wavelength $\lambda$ derived from the dispersion curves in (\textbf{B}) (modes $m=1,2$ shown). Mode $m=1$ forms a slow-wave branch with low phase velocity and short wavelength. Gray line: $\mathrm{A}_0$ mode of the homogeneous plate for comparison.}
    \label{fig:numerical-sims}
\end{figure*}

\subsection*{Resonator coupling produces slow waves in the tactile band} 

To facilitate the design of wave-based tactile displays with enhanced focal resolution, we numerically investigated metamaterial plates with arrays of local resonators. Iterative numerical simulation enabled us to engineer the dispersion relation of this system to support centimeter-scale focusing within the tactile frequency range. We simulated a family of systems comprising a thin acrylic plate ($h = \SI{1.5}{\milli\meter}$) coupled to a square lattice of cylindrical brass rod resonators ($\diameter = \SI{3.175}{\milli\meter}$, $h_\textrm{rod} = \SI{33.46}{\milli\meter}$, spacing $a = \SI{8}{\milli\meter}$; Fig.~\ref{fig:numerical-sims}A). Rod resonators were selected because they have been shown to efficiently couple to plate modes in other metamaterial systems~\cite{rupin2014experimental,williams2015theory,daunizeau2021phononic}. The resonator dimensions and lattice spacing were chosen to produce a slow-wave branch spanning the tactile frequency range (see Materials and Methods).

The resulting design yielded a fundamental dispersion branch ($m=1$) extending from \num{0} to \SI{558}{\hertz} (Fig.~\ref{fig:numerical-sims}B), with phase velocities below \SI{10}{\meter/\second} over this range (Fig.~\ref{fig:numerical-sims}C). Correspondingly, predicted wavelengths ranged from \num{1.1} to \SI{10}{\centi\meter} over \num{558} to \SI{75}{\hertz} (Fig.~\ref{fig:numerical-sims}D), supporting the generation of localized virtual pixels. These wavelengths were substantially smaller than those of the homogeneous plate; for example, at \SI{300}{\hertz}, the predicted wavelengths on the metamaterial and homogeneous plate were \num{3.0} and \SI{12.7}{\centi\meter}, respectively (see Supplementary Materials and Fig.~S1 for an analysis of the roles of mass loading and mode hybridization in wavelength reduction). Mode-shape analysis confirmed that this branch involves coupled plate flexion and lateral resonator bending, consistent with hybridization with the first resonator-associated mode at frequency $f_n=\omega_r/2\pi=\SI{781}{\hertz}$ (Fig.~\ref{fig:numerical-sims}B, inset). Although several branches fall within the tactile frequency range, wave propagation in the operating band is dominated by $m=1$, with higher-order branches contributing weakly or becoming relevant only at higher frequencies.

\subsection*{Metamaterial displays enable centimeter-scale virtual pixels}

\begin{figure*}[t!]
    \centering
    \includegraphics[scale=0.9]{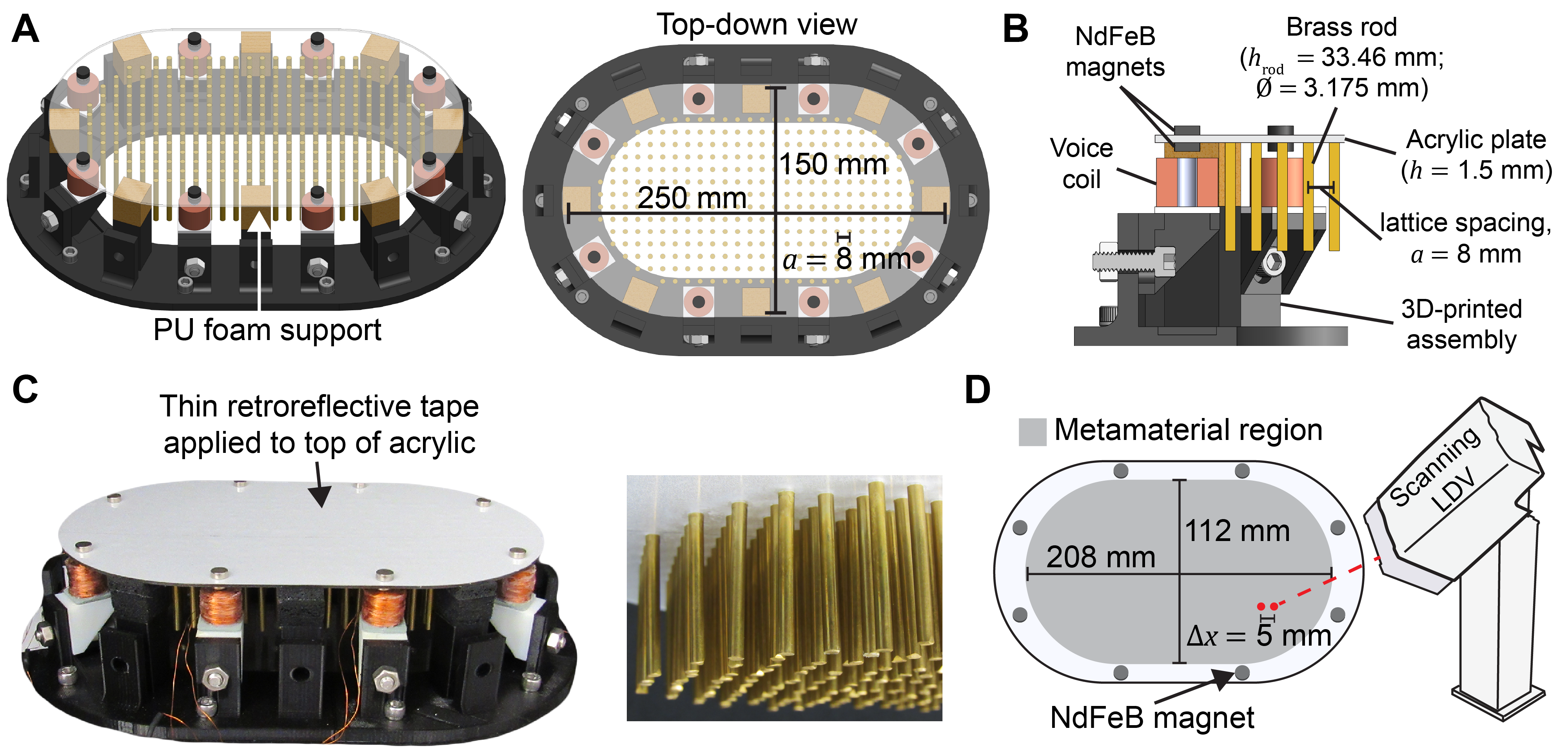}
    \caption{\textbf{Fig. 3. Metamaterial tactile display for spatially addressable tactile feedback.} (\textbf{A}) Device consisting of a \num{25} $\times$ \SI{15}{\centi\meter} thin acrylic plate (thickness: \SI{1.5}{\milli\meter}) coupled to more than 300 brass rods arranged in a square lattice (lattice constant $a=\SI{8}{\milli\meter}$). (\textbf{B}) Flexural plate waves are excited by voice coil actuators driving pairs of magnets affixed to the plate. (\textbf{C}) Fabricated metamaterial tactile display (left) and a close-up of rod resonators (right). Thin retroreflective tape was applied to the top of the acrylic plate after assembly to facilitate vibrometry measurements. (\textbf{D}) Scanning laser Doppler vibrometry was used to measure the spatially and temporally resolved surface velocity, $v(\mathbf{x},t)$, on a finely sampled grid ($\Delta x=\SI{5}{\milli\meter}$; 1225 measurement locations) during actuation.}
    \label{fig:device}
\end{figure*}

Our theoretical and numerical results informed the design of a physical metamaterial display demonstrating our approach (Fig.~\ref{fig:device}).  
In a series of focusing experiments, we show that centimeter-scale virtual pixels, with prescribed temporal waveforms, can be rendered at arbitrary surface locations using just eight actuators.

We fabricated a metamaterial tactile display in which flexural waves were excited and controlled by 8 peripherally located voice-coil actuators (Fig.~\ref{fig:device}A, B, C). These actuators applied in-plane forces to NdFeB magnets affixed to the plate. We captured spatially and temporally resolved optical vibrometry measurements of the normal plate surface velocity, $v(\mathbf{x},t)$, during excitation by each actuator (Fig.~\ref{fig:device}D). These measured impulse responses were used to analyze plate dynamics, evaluate driven responses under prescribed actuation, and compute the actuator driving signals for the behavioral experiments (see Materials and Methods).

\begin{figure*}[t!]
    \centering
    \includegraphics[scale=0.9]{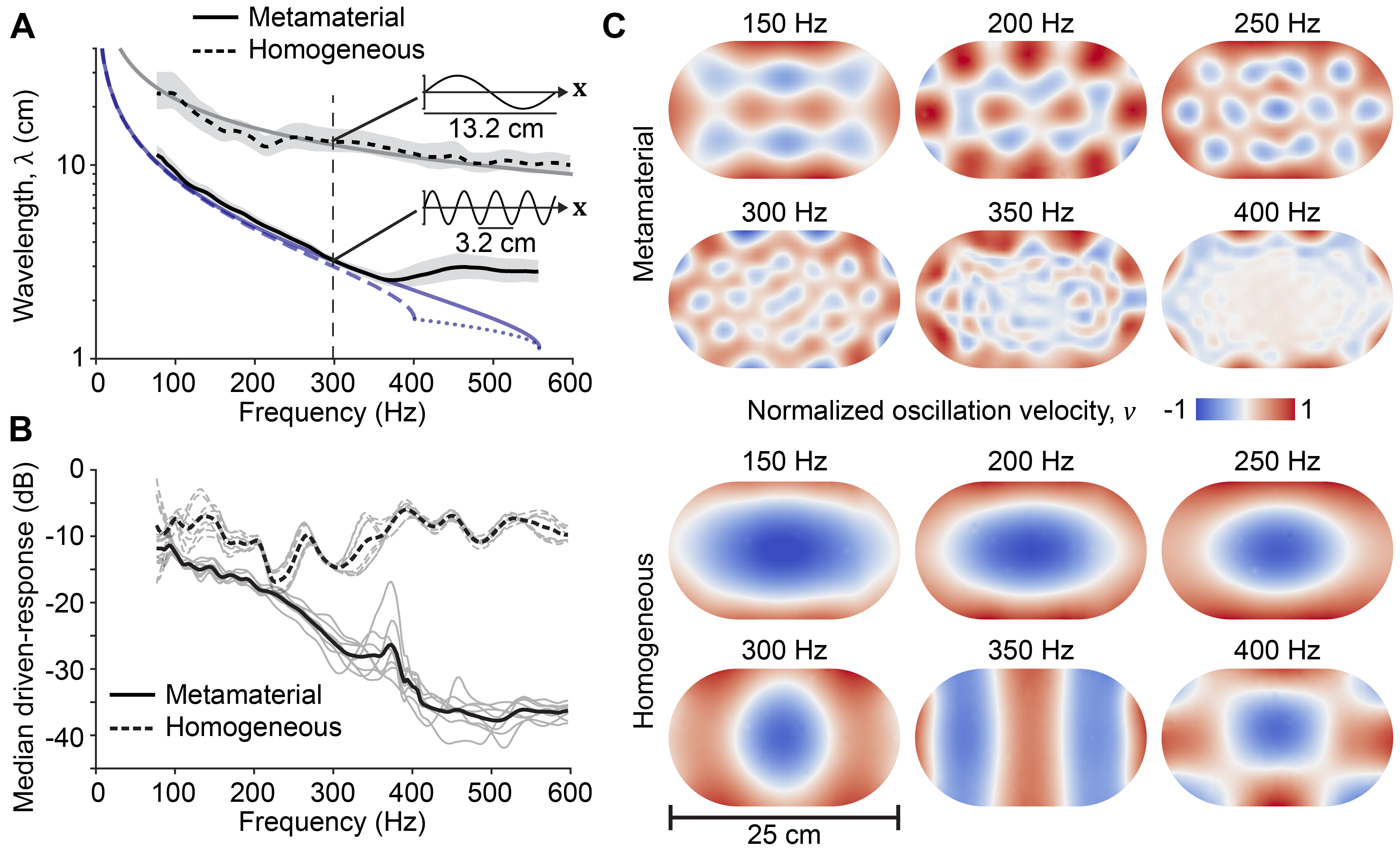}
    \caption{\textbf{Fig. 4. Measured slow-wave dispersion in the metamaterial tactile display.} (\textbf{A}) Experimentally measured wavelength $\lambda$ for the metamaterial plate (solid black line) and homogeneous plate (dashed black line). Black lines: mean of per-actuator median wavelengths; gray shading: mean of per-actuator interquartile bounds. Predictions from Fig.~\ref{fig:numerical-sims}D are overlaid: metamaterial mode $m=1$ (dark blue lines) and $\mathrm{A}_0$ mode of the homogeneous plate (gray line). (\textbf{B}) Median driven response for the metamaterial plate (black line) and homogeneous plate (black dashed line), normalized by the corresponding actuator responses (Fig.~S3) and averaged across actuators. Gray curves show per-actuator estimates. (\textbf{C}) Measured wave fields produced by sinusoidal driving signals between \num{150} and \SI{400}{\hertz}, shown at representative time frames for the metamaterial plate (top) and homogeneous plate (bottom). Red and blue indicate positive and negative instantaneous normal surface velocity, respectively.}
    \label{fig:plate-comparison}
\end{figure*}

Measurements confirmed that the metamaterial plate exhibited reduced wavelengths and lower phase velocities, in agreement with the numerical dispersion predictions. Mean wavelengths $\lambda(f)$ on the metamaterial plate ranged from \num{11.3} to \SI{2.5}{\centi\meter} over the frequency range of \num{75} to \SI{400}{\hertz} (Fig.~\ref{fig:plate-comparison}A). Over this range, wavelengths on the metamaterial plate ranged from \num{2.1} to \num{4.8} times smaller than those of a homogeneous plate (Fig.~S2). The median driven plate response generally decreased with increasing frequency (Fig.~\ref{fig:plate-comparison}B; referenced to the actuator response, Fig.~S3). This attenuation likely reflects a combination of impedance mismatch at the metamaterial boundary, intrinsic losses, increased resonator participation, and other loss-related mechanisms in the coupled plate--resonator system. These wavelength and attenuation measurements were corroborated by full-field observations of plate oscillations under sinusoidal excitation. The spatial wave patterns varied with excitation frequency, and the metamaterial plate exhibited shorter wavelengths than the homogeneous plate at each frequency (Fig.~\ref{fig:plate-comparison}C). At higher frequencies, vibration amplitudes within the resonator-covered region of the metamaterial plate also decreased, consistent with the decreasing driven response. Above approximately \SI{400}{\hertz}, this reduced response also made the dominant wavelength difficult to identify reliably from the measured wave fields (Fig.~\ref{fig:plate-comparison}A; Fig.~S4). Wave-focusing experiments were thus limited to \num{75}--\SI{400}{\hertz}.

\begin{figure*}[t!]
    \hspace*{-2mm}
    \includegraphics[scale=0.9]{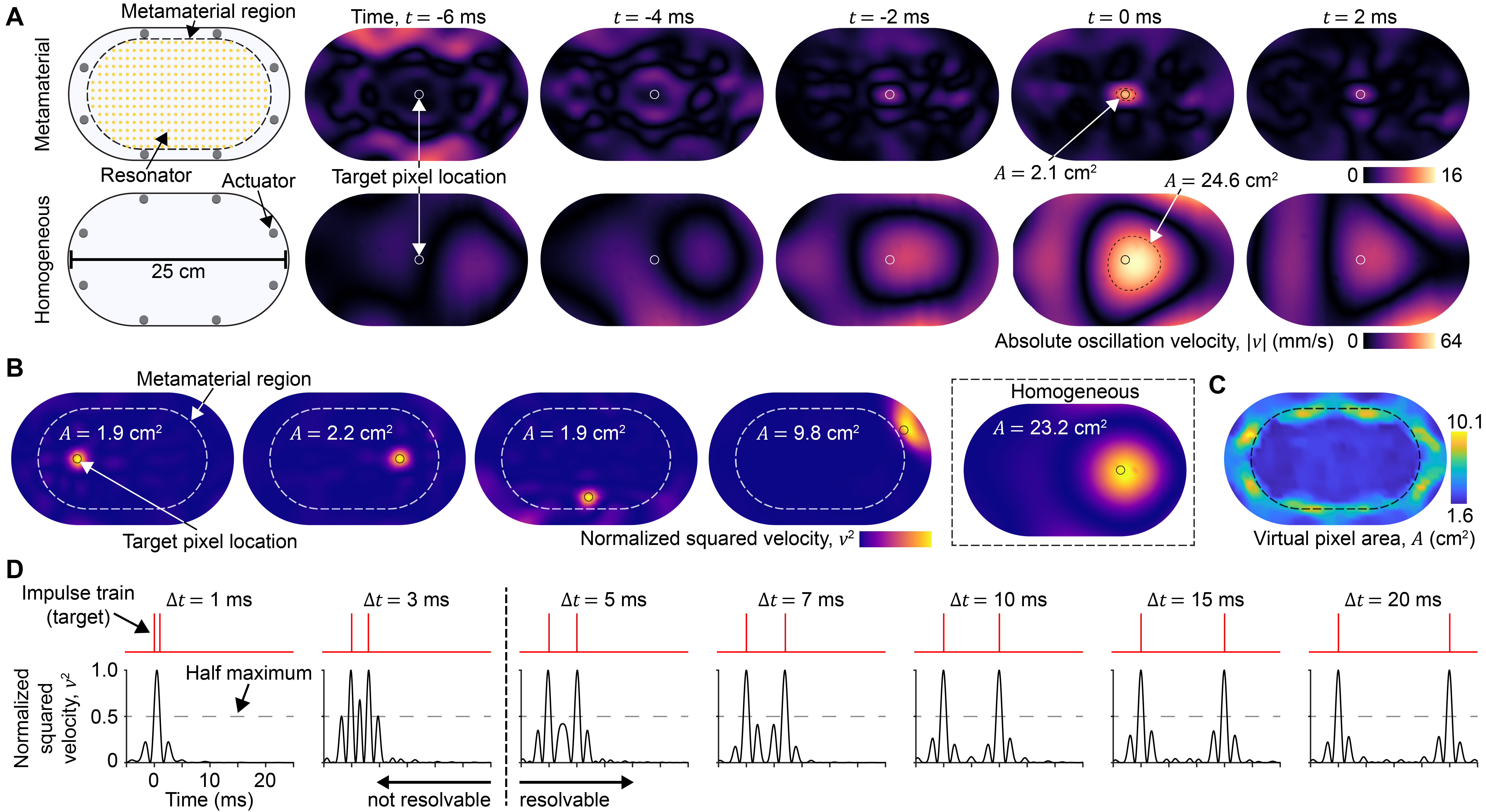}
    \caption{\textbf{Fig. 5. Metamaterial wave focusing for centimeter-scale tactile feedback: experimental measurements for single virtual pixels.} (\textbf{A}) Focusing causes plate waves to converge at the specified location, yielding virtual pixels (absolute plate velocity $|v(\mathbf{x},t)|$ shown; $t=0$ denotes the focal time, with negative times indicating instants before focus; circle: target pixel location). The metamaterial plate (top row) produces virtual pixels with area, $A$, an order of magnitude smaller than those on the homogeneous plate (bottom row). (\textbf{B}) Localized virtual pixels can be generated at arbitrary locations within the metamaterial display region (focusing frame at $t=0$ shown for different target locations; circle: target pixel location). Examples outside the resonator-covered region (plate periphery; fourth panel) and on the homogeneous plate (far right panel) are included for reference. (\textbf{C}) Spatial map of the virtual pixel area $A$, obtained by repeating the single-pixel focusing analysis across the metamaterial display (pixel area estimated at $t=0$). (\textbf{D}) Time-domain waveform at the pixel location when generating two bandlimited impulses separated by $\Delta t$ (pixel location: center of display). The wide operating bandwidth (\num{75}--\SI{400}{\hertz}) enables temporal impulses to be resolvable at separations as short as \SI{5}{\milli\second}.}
    \label{fig:focusing-singlepoint}
\end{figure*}

The shorter wavelengths on the metamaterial plate enabled wave focusing beyond the diffraction limit imposed by the homogeneous plate. In a series of focusing experiments using spatiotemporal inverse filtering (see Materials and Methods), we generated virtual pixels on the plate, achieving virtual pixel areas over an order of magnitude smaller than those achieved on the homogeneous plate~(Fig.~\ref{fig:focusing-singlepoint}A). The median virtual pixel area at focus time was \SI{2.74}{\centi\meter\squared} (Fig.~\ref{fig:focusing-singlepoint}B, C), comparable to the contact area of an adult index fingerpad under light loads during touch contact~\cite{van2015review}. The homogeneous plate did not consistently produce well-defined foci across all spatial locations, making it difficult to compute an analogous median virtual pixel area (Fig.~S5). Alternative focusing methods on the metamaterial plate, i.e., time-reversal focusing, produced similar virtual pixel areas (Fig.~S6). Spatiotemporal inverse filtering also allows us to prescribe the temporal waveform at the pixel location, including both transient and sustained vibrations. For example, we generated pairs of bandlimited impulses at a single pixel location with varying temporal separations (Fig.~\ref{fig:focusing-singlepoint}D). The wide operating bandwidth (\num{75}--\SI{400}{\hertz}) enables temporal impulses to be resolvable at separations as short as \SI{5}{\milli\second}.

\subsection*{Multi-pixel focusing is enhanced by metamaterials}

The preceding analyses demonstrated that the metamaterial plate supports shorter wavelengths, enabling the generation of highly localized virtual pixels. In a series of analyses of measured impulse responses and multi-pixel focusing experiments, we investigated how metamaterials enhance the generation of spatially addressable feedback with few actuated DOFs. 

\begin{figure*}[t!]
    \hspace*{-2mm}
    \includegraphics[scale=0.9]{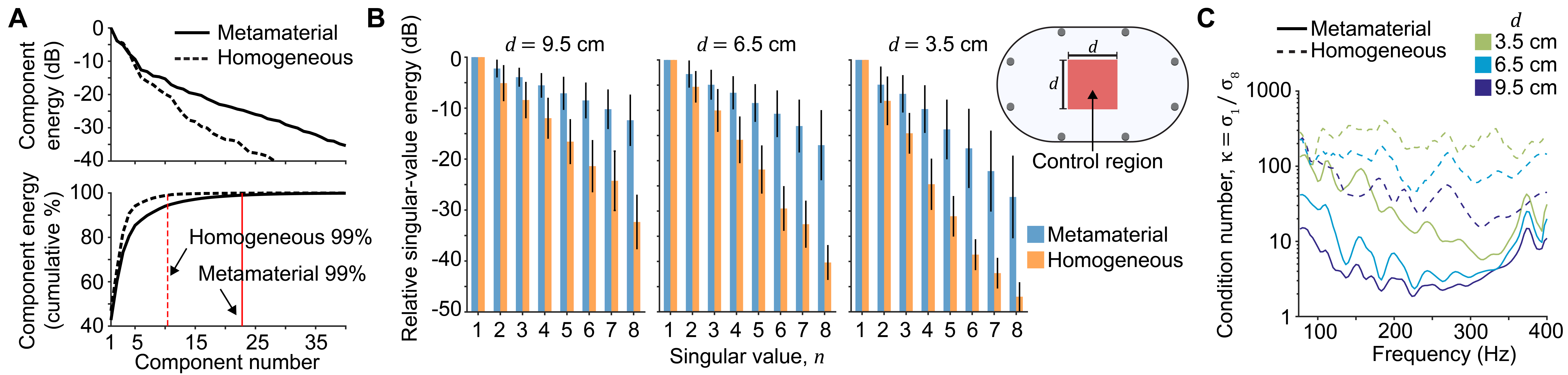}
    \caption{\textbf{Fig. 6. Dispersion engineering using locally resonant metamaterials increases spatial response dimensionality and input-output control authority.} (\textbf{A}) The metamaterial plate exhibits approximately a twofold increase in measured spatial response dimensionality relative to the homogeneous plate. Top: normalized POD component energy as a function of component number. Bottom: cumulative POD component energy as a percent of total energy (red lines: number of components required to capture \num{99}\% of response variance). (\textbf{B}) The metamaterial plate maintains higher actuator-addressable output strength across more singular directions than the homogeneous plate. Relative singular-value energy (band-averaged), $10\log_{10}(\sigma_i^2/\sigma_1^2)$, of the propagation matrix $\mathbf{G}_D(\omega)$ is shown for square domains of edge length $d=\num{9.5}$, \num{6.5}, and \SI{3.5}{\centi\meter}; error bars indicate standard deviation across frequencies from \num{75} to \SI{400}{\hertz}. (\textbf{C}) Condition number $\kappa=\sigma_1/\sigma_8$ of $\mathbf{G}_D(\omega)$ for different domain sizes on the metamaterial and homogeneous plates. Lower values indicate a smaller separation between the strongest and weakest actuator-addressable response directions, supporting more stable multi-pixel wave field generation.}
    \label{fig:plate-controllability}
\end{figure*}

Simultaneous multi-pixel generation requires access to multiple independently addressable spatial response patterns on the display surface. We therefore first quantified the spatial dimensionality of the measured impulse-response fields using proper orthogonal decomposition (POD). The metamaterial plate required 23 spatial POD components to capture 99\% of the response variance, compared with 11 components for the homogeneous plate, indicating an approximately twofold increase in the measured spatial response dimensionality (Fig.~\ref{fig:plate-controllability}A; Fig.~S7).

This increased spatial response dimensionality expands the set of wave field configurations that can be addressed through the actuator array, thereby increasing input-output authority. We evaluated this authority over localized square regions $D$ on the plate surface by constructing frequency-dependent propagation matrices from impulse responses measured within each domain (edge length $d$; Fig.~\ref{fig:plate-controllability}B, right). This yielded matrices $\mathbf{G}_D(\omega)$ of size $M_D \times 8$, where $M_D=\num{361}$, \num{169}, and \num{49} measured locations for $d=\num{9.5}$, \num{6.5}, and \SI{3.5}{\centi\meter}, respectively. These matrices map the eight frequency-domain actuator inputs, $\mathbf{a}(\omega)$, to the plate response within each domain, $\mathbf{v}_D(\omega)$:
\begin{equation}
    \mathbf{v}_D(\omega) = \mathbf{G}_D(\omega) \mathbf{a}(\omega).
    \label{eq:metaplate-greensfunction-solution}
\end{equation}
We decomposed $\mathbf{G}_D(\omega)$ using singular value decomposition, where each singular value quantifies how strongly an actuator drive pattern excites its corresponding spatial response pattern.

The metamaterial plate maintained higher relative singular-value energy (band-averaged) across more singular directions than the homogeneous plate, and this difference became more pronounced as domain size decreased (Fig.~\ref{fig:plate-controllability}B; Fig.~S8). These results indicate that the metamaterial plate preserves more actuator-addressable output directions when control is restricted to smaller regions of the display, supporting independent generation of multiple nearby virtual pixels. We summarized this input-output authority using the band-averaged effective rank of $\mathbf{G}_D(\omega)$, defined using a \SI{-30}{\decibel} singular-value-energy threshold. For the two larger domains ($d=\num{9.5}$ and \SI{6.5}{\centi\meter}), the metamaterial plate approached the maximum effective rank permitted by the eight-actuator array. For the smallest region examined ($d=\SI{3.5}{\centi\meter}$), the metamaterial plate retained an average effective rank of \num{7.4}, compared to \num{4.3} for the homogeneous plate.

This enhanced input-output authority aids multi-pixel generation by allowing target wave fields to be generated with lower actuator drive amplitudes and reduced sensitivity to noise, calibration errors, and perturbations to the actuator-to-surface response. We assessed this stability using the condition number $\kappa=\sigma_1/\sigma_8$ of $\mathbf{G}_D(\omega)$, computed from all eight singular values (Fig.~\ref{fig:plate-controllability}C). The metamaterial plate exhibited modest condition numbers across most frequencies and domains ($\bar{\kappa}=\num{5.0}$, \num{10.1}, and \num{35.9} for $d=\num{9.5}$, \num{6.5}, and \SI{3.5}{\centi\meter}, respectively), substantially below the corresponding values for the homogeneous plate ($\bar{\kappa}=\num{51.5}$, \num{110.6}, and \num{236.0}). The lower condition numbers measured for the metamaterial plate within these localized regions indicate that the underlying actuator-to-surface maps used to compute actuator signals for multi-pixel focusing are better conditioned, reducing amplification of weak response directions and supporting more robust wave field generation.

\begin{figure*}[t!]
    \hspace*{-1mm}   
    \includegraphics[scale=0.9]{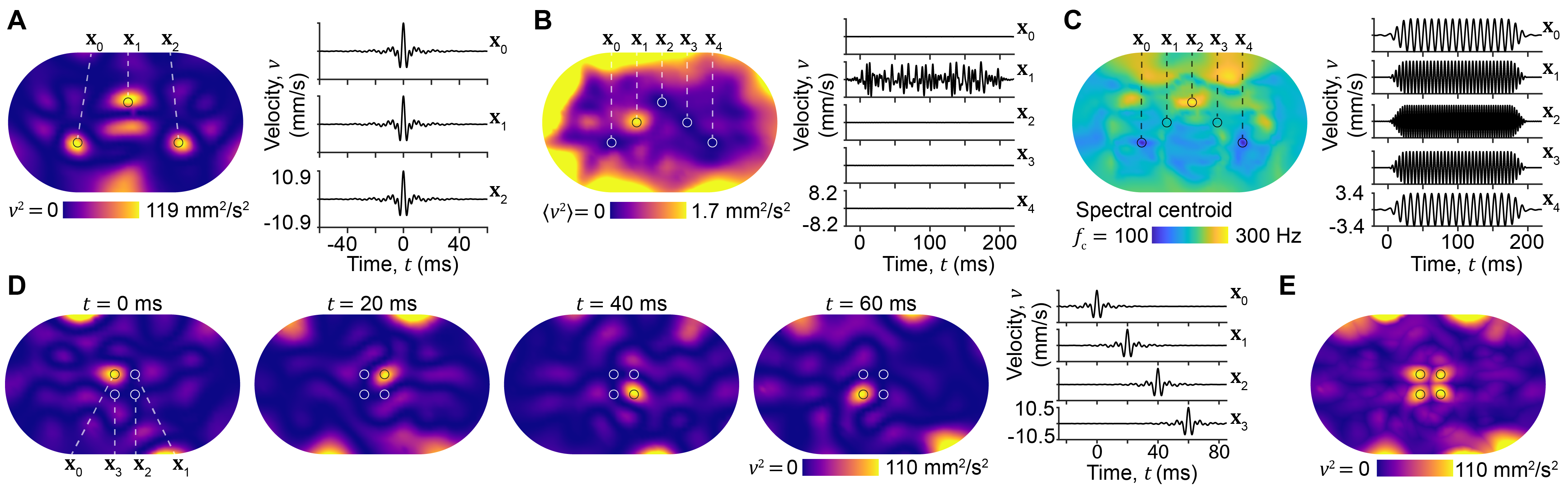}
    \caption{\textbf{Fig. 7. Virtual pixels for multi-point tactile feedback.} (\textbf{A}) Multiple virtual pixels can be activated simultaneously with precise temporal control (target waveform: bandlimited impulse). Left: squared wave field velocity $v(\mathbf{x},t)^2$ at focus time $t=0$. Right: time-domain velocity at pixel locations (peak displacement: \SI{9.0}{\micro\metre}). (\textbf{B}) Virtual pixels can also be assigned null targets to suppress unintended vibrations at selected locations. Broadband noise is delivered to one virtual pixel while the responses at four others are nulled. Left: time-averaged squared velocity, $\langle v(\mathbf{x},t)^2 \rangle_t$. Right: time-domain velocity at pixel locations (peak displacement: \SI{9.7}{\micro\metre}). (\textbf{C}) Virtual pixels can be driven with distinct temporal signals, such as sinusoids of different frequencies, to generate complex multi-location tactile patterns. Left: spectral centroid, $f_c(\mathbf{x})$, characterizing the local frequency content of surface vibrations, computed from the temporal Fourier transform of $v(\mathbf{x},t)$. Right: time-domain velocity at pixel locations (peak displacements at locations $\mathbf{x}_0$--$\mathbf{x}_4$, respectively: \num{5.5}, \num{2.7}, \num{1.8}, \num{2.7}, and \SI{5.5}{\micro\metre}). (\textbf{D}) Virtual pixels can be sequentially activated to generate dynamic tactile patterns. Left: squared velocity $v(\mathbf{x},t)^2$ at focusing frames $t_f=0$, \num{20}, \num{40}, and \SI{60}{\milli\second}. Right: time-domain velocity at pixel locations (peak displacement: \SI{8.2}{\micro\metre}). (\textbf{E}) Maximum squared velocity across the focusing frames shown in (\textbf{D}), $\max_{t\in t_f} v(\mathbf{x},t)^2$.}
    \label{fig:focusing-kitchensink}
\end{figure*}

Together, the increased spatial response dimensionality and input-output authority enable the generation of multiple localized virtual pixels. We generated independent, computationally specified tactile signals at multiple pixel locations using spatiotemporal inverse filtering; each pixel is a localized excitation of the plate with an independent waveform. Virtual pixels were first generated simultaneously at three surface locations by prescribing a bandlimited impulse waveform at each location (Fig.~\ref{fig:focusing-kitchensink}A). The resulting pixels were well localized, with areas comparable to those of the single-pixel experiments (Fig.~\ref{fig:focusing-singlepoint}B, C), and exhibited peak velocities exceeding \SI{10}{\milli\meter/\second} and peak displacements of \SI{9}{\micro\metre}, significantly above perceptual thresholds~\cite{verrillo1963effect,verrillo1971vibrotactile}. Broadband noise was then generated at one pixel location while vibrations were actively suppressed at four others (Fig.~\ref{fig:focusing-kitchensink}B). Finally, sinusoidal oscillations of varying frequency (\num{100} to \SI{300}{\hertz}) were generated at five pixel locations (Fig.~\ref{fig:focusing-kitchensink}C). This programmable control over plate oscillations at multiple spatial locations can be used to design a myriad of different haptic feedback patterns. For example, patterns can also be sequenced over time, such as delivering feedback in sequence along a square trajectory (Fig.~\ref{fig:focusing-kitchensink}D, E).

\subsection*{Virtual tactile pixels present perceptible dynamic tactile patterns}

Perception of these multi-pixel patterns was then evaluated across four behavioral experiments. These experiments employed both transient and sustained vibrations to assess virtual-pixel localization, active exploration, motion-direction perception, and simultaneous multi-finger discrimination (Fig. 8; Figs. S9–S12; see Supplementary Materials for stimulus details). Ten participants with no prior training took part in the experiments, although one did not complete experiment 4.

\begin{figure*}[t!]
    \centering
    \includegraphics[scale=0.9]{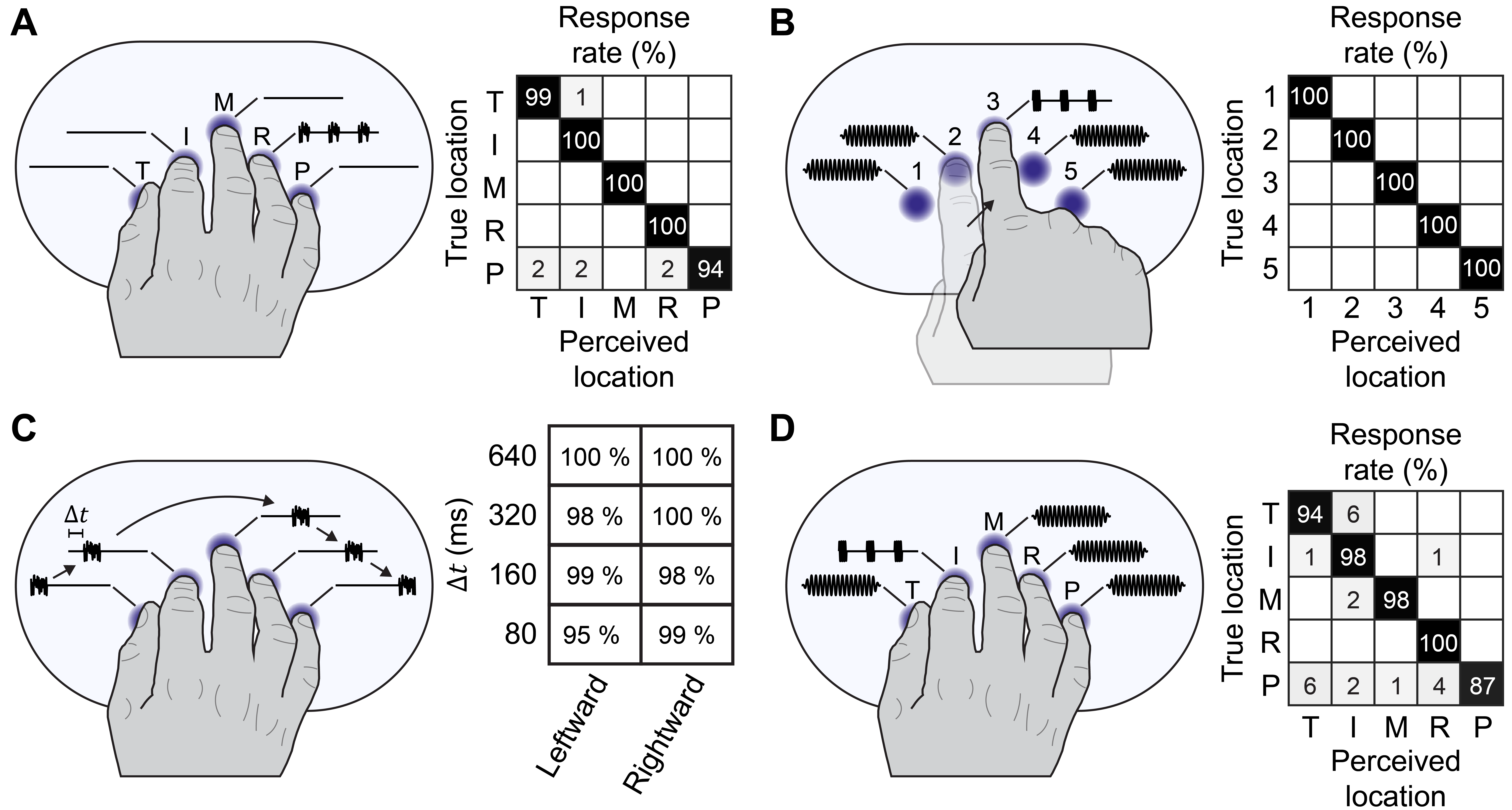}
    \caption{\textbf{Fig. 8. Perception of multiple virtual pixels.} (\textbf{A}) Experiment 1: Virtual pixel localization. Participants identified the active virtual pixel (left). Mean response accuracy was \num{98.6}\% (right; chance accuracy: \num{20}\%). (\textbf{B}) Experiment 2: Virtual pixel exploration. Participants explored five virtual pixels and identified the pixel with a distinct vibration pattern (left). Participants correctly identified the target pixel on every trial (right; chance accuracy: \num{20}\%). (\textbf{C}) Experiment 3: Virtual pixel sequences. Participants identified the direction of sequentially activated virtual pixels (left). The interval between activations, $\Delta t$, was varied, with smaller $\Delta t$ corresponding to faster sequential activation. Mean response accuracy was \num{98.6}\% (right; chance accuracy: \num{50}\%). (\textbf{D}) Experiment 4: Simultaneous multi-pixel patterns. Tactile signals were presented simultaneously at five virtual pixels, and participants identified the pixel with a distinct vibration pattern (left). Mean response accuracy was \num{95.3}\% (right; chance accuracy: \num{20}\%).}
    \label{fig:focusing-perception}
\end{figure*}

Experiment 1 assessed the perception of localized bursts of haptic feedback delivered at different virtual pixels on the display (Fig.~\ref{fig:focusing-perception}A, Fig.~S9). Participants were asked to identify which of five pixel locations, corresponding to locations under each finger of their right hand, was activated. Participants correctly identified the active pixel in \num{98.6}\% of trials (5 stimuli $\times$ 10 repetitions $\times$ 10 participants; 493/500 total trials correct). In experiment 2, participants freely explored five virtual pixels with their index finger and identified the pixel presenting a distinct haptic pattern (``odd-one-out'' paradigm; Fig.~\ref{fig:focusing-perception}B, Fig.~S10). Participants correctly identified the odd-one-out pixel in \num{100}\% of trials (10 stimuli $\times$ 5 repetitions $\times$ 10 participants; 500/500 total trials correct).

Experiment 3 assessed participants' perception of motion patterns generated by sequential activation of virtual pixels beneath the fingertips of their right hand. Participants were asked to identify the direction of sequential pixel activation as either leftward or rightward across their fingers, similar to prior wave-mediated haptic display studies~\cite{reardon2023rendering} (Fig.~\ref{fig:focusing-perception}C, Fig.~S11). We varied the time interval, $\Delta t$, between sequential pixel activations, with smaller $\Delta t$ corresponding to faster motion. Participants identified the correct direction in \num{98.6}\% of trials (4 motion speeds $\times$ 2 directions $\times$ 10 repetitions $\times$ 10 participants; 789/800 total trials correct).

In a final experiment, we investigated whether participants could localize a distinct haptic pattern when multiple virtual pixels were simultaneously activated and contacted. Participants received feedback at five pixels, corresponding to locations under each finger of their right hand, with one pixel presenting a feedback pattern that differed from the other four (``odd-one-out'' paradigm; similar to experiment 2, except all fingers contacted the display; Fig.~\ref{fig:focusing-perception}D, Fig.~S12). Participants correctly identified the odd-one-out pixel in \num{95.3}\% of trials (5 stimuli $\times$ 10 repetitions $\times$ 9 participants; 429/450 total trials correct).

\section*{DISCUSSION}
This research shows how localized tactile feedback across an extended surface can be generated by engineering wave dynamics using metamaterial principles, enabling remote actuation from a sparse peripheral actuator array rather than a dense array of physical tactile pixels. Coupling the surface to an array of local resonators modified the dispersion relation and reduced the wavelengths that constrain focusing. With this engineered medium, eight peripheral actuators were sufficient to focus energy at arbitrary locations on the surface, yielding multiple independently addressable virtual pixels. Our findings enable dynamic tactile surfaces whose spatial resolution is facilitated by material design, mechanical architecture, and computational control.

Using physics-based analysis and numerical optimization, we identified a locally resonant plate geometry with a slow-wave branch in the tactile frequency range (here, \num{75}--\SI{400}{\hertz}). Full-field vibrometry confirmed that the fabricated metamaterial plate reduced plate wavelengths by factors of approximately 2 to 5 relative to a homogeneous plate. We then employed computational wave-field control to generate virtual pixels with a median focal area of \SI{2.74}{\centi\meter\squared}, an order-of-magnitude improvement over a homogeneous plate. Behavioral experiments showed these wave fields were perceptually salient: participants could readily localize, explore, discriminate, and perceive spatiotemporal tactile patterns produced by the display.

Our results further show that virtual tactile pixels are constrained by the spatial response patterns supported by the surface. By reshaping these response patterns through dispersion engineering, the architecture enabled greater input-output authority and finer focusing and control than would be feasible in a conventional surface. These results are consistent with recent work applying spatiotemporal inverse filtering to homogeneous displays, which demonstrated that strong spatial correlations between propagation responses at nearby control locations can lead to poorly conditioned inverse problems and increased waveform reconstruction error~\cite{hudin2026spatio}. By engineering the metamaterial to support higher wavenumbers at frequencies relevant to vibrotactile perception, our approach enhances the ability to independently control nearby locations, as reflected in the increased spatial response dimensionality and improved propagation-matrix conditioning (Fig. 6). This research also extends the role of locally resonant metamaterials, which are often studied for passive effects such as attenuation, filtering, isolation, wave guiding, or energy localization. Here, the engineered dispersion relation is used generatively, enabling an actively driven surface to deliver mechanical energy to selected locations and at spatial scales relevant to multi-digit touch interactions.

The improvement in spatial localization is accompanied by tradeoffs in vibration amplitude, excitation efficiency, and operating bandwidth. For the metamaterial design investigated here, the driven response of the plate decreases at higher frequencies (Fig.~\ref{fig:plate-comparison}B), motivating our use of a \num{75}--\SI{400}{\hertz} operating band that includes the range of greatest vibrotactile sensitivity. Under identical peak actuator-voltage constraints, the metamaterial plate produced approximately one-quarter of the peak focal velocity measured on the homogeneous plate in a representative focusing experiment (Fig.~\ref{fig:focusing-singlepoint}A). Nevertheless, the resulting vibrations remained well above reported perceptual thresholds and supported the tactile functions demonstrated in our behavioral experiments. The reduced driven response likely reflects several contributing mechanisms, including impedance mismatch between the unmodified plate periphery (where the actuators are located) and the resonator-covered region~\cite{lott2019effective}, together with frequency-dependent resonator dynamics and mechanical losses. Future designs could explore actuation within the resonator-covered region or modifications to the transition between the two regions as potential means of improving excitation efficiency and extending the usable operating bandwidth, as well as alternative resonator architectures that target different operating bands. These considerations highlight the need to balance spatial localization, bandwidth, and vibration amplitude when designing future metamaterial tactile displays.

Our implementation of wave-field control relies on measured actuator-to-surface propagation responses, which can change with contact state, actuator mounting, fabrication variation, or long-term drift. Such changes may introduce waveform reconstruction errors and motivate recalibration, contact sensing, or adaptive identification. However, our behavioral experiments demonstrated that participants could reliably localize and discriminate tactile patterns during single- and multi-finger contact, including simultaneous discrimination at five locations with 95.3\% accuracy, despite applied finger forces not being controlled. Thus, across ten participants and more than 2,000 experimental trials, the delivery of localized tactile feedback remained robust to variations in how users contacted the display. Recent work on an OLED-based tactile display reported relatively small deviations in measured propagation responses during finger contact (median deviation $<$12\% for up to 10 fingers), attributed in part to the high mechanical impedance of the display relative to the finger~\cite{hudin2026spatio}. While these findings are encouraging, the effects of finger loading on wave propagation and waveform reconstruction in our metamaterial display remain to be characterized.

While we demonstrate one embodiment, our approach combining dispersion engineering and computational wave-field control provides a basis for a range of display designs, including configurations with different numbers of actuators. Future designs may prioritize different practical considerations, including display thickness, manufacturability, or tunability. Lower-profile resonators and folded or planar inclusions could reduce display thickness and facilitate integrated fabrication, improving manufacturability. Replacing the mechanical resonators with piezoelectric-inductive elements could also reduce display thickness and enable tuning of the dispersion characteristics, albeit with significantly increased electronic complexity. Tunable mechanical or electromagnetic resonators offer alternative means of adjusting the display's dispersion characteristics. One possibility for integrating visual feedback is to bond a flexible OLED display to the top surface of the acrylic plate. Complementary actuation mechanisms could, in principle, extend the range of tactile feedback to include low-frequency vibrations and quasi-static surface deformation, but practical methods for delivering localized feedback of this kind across large display areas remain limited. Alternative metamaterial designs could also target shorter wavelengths to achieve finer spatial resolution of vibrotactile feedback, potentially approaching the millimeter scale, although these improvements would need to be balanced against operating bandwidth, attenuation, and excitation efficiency.

Integrating touch sensing could enable interactive applications in which tactile feedback responds to finger position and movement. Continuous feedback during finger movement would require updating the actuator driving signals in real time as contact locations change. Existing overlap-and-add methods for real-time wave-field control could be extended to the metamaterial display presented here~\cite{hudin2026spatio}, although their effectiveness remains to be investigated. The sparse actuation approach demonstrated here also offers opportunities for larger-area displays, including those supporting multiple users. The present implementation demonstrated independently addressable tactile feedback across a surface exceeding 300 cm² using eight peripheral actuators, but extending this capability to larger areas and more simultaneously controlled locations will depend on wave propagation, attenuation, and the number and placement of actuators.

More broadly, this work suggests a different scaling principle for tactile displays. Rather than a simple association of one actuator to each surface location, it proposes tactile surfaces whose geometry and dynamics abstract this relationship, making sparse actuation spatially expressive. In such systems, computation determines the location and parameters of what will be felt, while the engineered material architecture determines the wave process that translates it into localized mechanical signals.

\section*{MATERIALS AND METHODS}

\subsection*{Numerical dispersion experiments}

Flexural wave dispersion was characterized across candidate metamaterial designs using finite element eigenfrequency analysis with Bloch periodic boundary conditions (COMSOL Multiphysics). Resonator configurations, materials, and dimensions were initially evaluated to identify designs capable of supporting slow waves within the tactile frequency range. Following the selection of an architecture consisting of a thin acrylic plate coupled to cylindrical brass-rod resonators, we varied resonator dimensions and lattice spacing to target a fundamental dispersion branch ($m=1$) with phase velocities $c_{\mathrm{meta}} < \SI{10}{\meter/\second}$, particularly within the \num{200}--\SI{300}{\hertz} range of high vibrotactile sensitivity, while seeking to position interactions with higher-order dispersion branches well above this range. The design was further refined under practical constraints on fabrication and static plate loading. Although local resonators need not form a lattice to modify wave dispersion, the lattice structure enables efficient design and analysis using Bloch eigenvalue methods. Additional details of the numerical experiments, including the material parameters, procedure for constructing the band diagram, and rationale behind the selected design, are provided in the Supplementary Materials.

The dispersion behavior of an identical homogeneous plate (i.e., without resonators) was also computed using our numerical procedure. These results were compared against the analytical solution for the flexural phase velocity~\cite{darabi2018broadband} for validation (Fig.~S13): 
\begin{equation}
    c(\omega)^4 = \omega^2 \frac{Eh^2}{12(1-\nu^2)\rho} .
\end{equation}
At low frequencies, these flexural waves travel with relatively slow phase velocities. However, for a thin acrylic plate ($E = \SI{3.2}{\giga\pascal}$, $h = \SI{1.5}{\milli\meter}$, $\nu = \num{0.37}$, $\rho = \SI{1180}{\kilo\gram/\meter\cubed}$), wavelengths ($\lambda = 2\pi c/\omega$) still range from \num{8} to \SI{50}{\centi\meter} over the frequency range of \num{800} to \SI{20}{\hertz}.

\subsection*{Tactile display fabrication}
The metamaterial and homogeneous plates were each composed of a thin acrylic plate (thickness: \SI{1.5}{\milli\meter}) with a Bunimovich stadium shape (parameters: $l = \SI{10}{\centi\meter}$, $r = \SI{7.5}{\centi\meter}$). An ergodic stadium design was used to reduce the effects of degenerate plate modes. To construct the metamaterial plate, brass rods (average height: \SI{33.46}{\milli\meter}, diameter: \SI{3.175}{\milli\meter}; Fig.~S14) were press-fit into laser-cut holes in the acrylic (square lattice; $a$ = \SI{8}{\milli\meter}).

The metamaterial and homogeneous plates were supported by eight foam support pillars with square cross-sections (\num{1.8} $\times$ \SI{1.8}{\centi\meter}) to provide compliant support that facilitated local plate flexion while reducing rigid-support constraints and losses. The plates were driven by an array of eight custom electromagnetic actuators distributed along the periphery. These actuators produced magnetic fields that exerted in-plane forces on NdFeB magnets affixed to the plate, driving high-amplitude plate waves. Actuators were controlled using commodity audio hardware (24-channel digital-to-analog converter: 24Ao, MOTU, USA; 12-channel audio amplifier: MA1240a, Dayton Audio, USA). Additional fabrication details are available in the Supplementary Materials.

\subsection*{Vibrometry measurements of tactile displays}
Spatially and temporally resolved surface oscillations $v(\mathbf{x},t)$ on the metamaterial and homogeneous plates were captured via scanning laser Doppler vibrometry (Model 8330, Ometron Ltd., UK) and a data acquisition unit (USB-6211, National Instruments, USA; \SI{24}{\kilo\hertz} sampling rate). Retroreflective tape (Reflective Solid Floor Marking Tape, McMaster-Carr, USA) was applied to the top surface of each plate to improve laser reflectivity. Green's functions $g_i(\mathbf{x},t)$, referred to here interchangeably as impulse responses, were captured from each actuator to each of 1225 locations on a uniform grid ($\Delta x = \SI{5}{\milli\meter}$; Fig.~\ref{fig:device}D) using \SI{3}{\second} linear sinusoidal sweep test signals (\num{62}--\SI{625}{\hertz}, 5 repetitions) and subsequent deconvolution. These Green's functions, formally, $g_i(\mathbf{x},t; \mathbf{x}',t')$ describe the response of the plate at location $\mathbf{x}$ and time $t$ to unit impulsive forces delivered by actuator $i$ at location $\mathbf{x}'$ and time $t'$. For convenience, $\mathbf{x}'$ is dropped, and $t' = 0$ is assumed, while $i$ is retained to denote an actuator with the associated location $\mathbf{x}'_i$. The entire measurement process took approximately 30 hours per plate. The ensemble of Green's functions encodes the linear physics of the actuator--plate system~\cite{hespanha2018linear}, enabling data-driven computational experiments and in-depth analyses of wave field oscillations. This methodology was validated in prior research~\cite{reardon2023rendering}.

\subsection*{Data-driven wave-field control}
The measured Green's functions, $g_i(\mathbf{x},t)$, from each actuator to each plate location were used for analyses and to compute the plate oscillation velocity, $v(\mathbf{x},t)$, in response to prescribed actuator driving signals, $a_i(t)$, in a series of \textit{in-silico} experiments. The linear wave field response in the normal direction can be described as
\begin{equation}
    v(\mathbf{x},t) = \sum_{i=1}^N g_i(\mathbf{x},t) * a_i(t)    
\end{equation} 
where $N$ is the number of actuators and $*$ denotes time-domain convolution. This was efficiently computed in the frequency domain:
\begin{equation}
    \mathbf{v}(\omega) = \mathbf{G}(\omega) \mathbf{a}(\omega)
    \label{eq:wavefield-solution}
\end{equation}
where $\mathbf{G}(\omega)$ is the Fourier transform of the Green's functions for all measured locations and actuators ($M\times N$ propagation matrix), $\mathbf{a}(\omega)$ is a vector of the $N$ actuator driving signals ($N\times1$), and $\mathbf{v}(\omega)$ is the frequency-domain wave field velocity ($M\times1$) at all measured locations. Time-domain wave field responses were then recovered via an inverse Fourier transform.

Virtual pixels in the single-pixel, multi-pixel, and behavioral experiments (Figs.~\ref{fig:focusing-singlepoint}, \ref{fig:focusing-kitchensink}, \ref{fig:focusing-perception}; see Supplementary Materials) were generated by specifying desired velocity signals, $\hat{v}(\mathbf{x}_c,t)$, at $C$ pixel locations and solving for the actuator driving signals. The target velocity signals were converted to frequency-domain target vectors $\hat{\mathbf{v}}_C(\omega)$ ($C\times 1$), and actuator signals were obtained by minimizing a least-squares cost function (i.e., spatiotemporal inverse filtering~\cite{tanter2001optimal}):
\begin{equation}
    \min_{\mathbf{a}(\omega)} ||\mathbf{G}_C(\omega)\mathbf{a}(\omega) - \hat{\mathbf{v}}_C(\omega)||_2^2 + ||\Gamma(\omega) \mathbf{a}(\omega)||_2^2,
\end{equation}
where $\mathbf{G}_C(\omega)$ ($C\times 8$) is the propagation matrix describing transmission from the eight actuators to the $C$ pixel locations, $\Gamma(\omega) = \alpha(\omega) \mathbf{I}$ is a Tikhonov matrix with regularization parameter $\alpha(\omega)$, and $||.||_2^2$ is the squared $\ell^2$-norm. Minimizing this function yields the regularized least-squares solution:
\begin{equation}
    \mathbf{a}^*(\omega) = (\mathbf{G}_C(\omega)^\dagger \mathbf{G}_C(\omega) + \Gamma(\omega)^\dagger \Gamma(\omega))^{-1}\mathbf{G}_C(\omega)^\dagger\hat{\mathbf{v}}_C(\omega)
    \label{eq:wavefield-control}
\end{equation}
where $^\dagger$ is the conjugate transpose. Additional details on the regularization procedure are provided in the Supplementary Materials. The actuator driving signals were computed at each frequency bin and subsequently converted to the time domain using an inverse Fourier transform. In this work, we investigated a wide range of target velocity signals, $\hat{v}(\mathbf{x}_c,t)$, with $C$ ranging from 1 to 5 (see next section for specific experimental details). 

\subsection*{Wave field analysis}

Plate wavelengths were estimated from the spatial Fourier transforms $g_i(\mathbf{k},\omega)$ of the measured Green's functions. At each temporal frequency, the complex-valued spatial response was multiplied by a two-dimensional Tukey window (taper fraction 0.3) prior to the spatial Fourier transform. For frequencies above \SI{280}{\hertz}, a Laplacian-of-Gaussian spatial filter was also applied to improve identification of shorter-wavelength features. The dominant peak in the magnitude of the resulting spatial Fourier spectrum was then identified for each actuator, and the surrounding full-width-at-half-maximum contour was computed. The median and interquartile bounds of the wavenumber magnitudes $|\mathbf{k}|$ within this contour were calculated and subsequently converted to wavelength ($\lambda = 2\pi / |\mathbf{k}|$). The per-actuator median wavelengths and interquartile bounds were then averaged across the eight actuators (Fig.~\ref{fig:plate-comparison}A). Wavelength-reduction factors were calculated for each actuator as the ratio of median wavelengths on the homogeneous and metamaterial plates, and subsequently averaged across actuators (Fig.~S2). The driven plate response was characterized by computing the median frequency response across measured surface locations for each actuator and then averaging across actuators (Fig.~\ref{fig:plate-comparison}B). Separately, the attenuation time constants of both plates were estimated from the temporal decay of their measured Green's functions (see Supplementary Materials; Fig.~S15).

Plate response patterns (Fig.~\ref{fig:plate-comparison}C) were first visualized using sinusoidal driving signals of varying frequency, $a_i(t)=\sin(2\pi f t)$. Finite-plate response dimensionality was later quantified using proper orthogonal decomposition (POD) of the measured Green's functions gathered across the plate and summed over actuators, $h(\mathbf{x},t)=\sum_{i=1}^N g_i(\mathbf{x},t)$. Responses from all 1225 measured surface locations were assembled into a matrix $\mathbf{H}$ and decomposed as $\mathbf{H}=\mathbf{U}\boldsymbol{\Sigma}\mathbf{V}^\top$. The columns of $\mathbf{U}$ define spatial POD components, the squared singular values in $\boldsymbol{\Sigma}$ quantify the response variance captured by each component, and the columns of $\mathbf{V}$ define the corresponding temporal coefficients (Fig.~\ref{fig:plate-controllability}A, Fig.~S7). This analysis does not identify all dynamic modes of the plate, but provides a compact summary of the coherent spatial response structures excited when driving the actuator array in phase.

Input-output authority was assessed using singular value decomposition (SVD) of the frequency-dependent propagation matrices $\mathbf{G}(\omega)$ (Fig.~\ref{fig:plate-controllability}B; Fig.~S8). For each square domain $D$ of edge length $d$, $\mathbf{G}(\omega)$ was restricted to the $M_D$ measured locations within that domain, yielding a domain-specific matrix, $\mathbf{G}_D(\omega)$ ($M_D \times 8$), describing transmission from the eight actuators to each measured point in the domain $D$. At each frequency, the SVD of $\mathbf{G}_D(\omega)$ returned $8$ singular values, each quantifying how strongly an actuator drive pattern excites its corresponding spatial response pattern within $D$. Relative singular-value energy was computed as $\sigma_{i,\mathrm{dB}}^2 = 10\log_{10}(\sigma_i^2/\sigma_1^2)$. The effective rank of $\mathbf{G}_D(\omega)$ was defined as the number of singular-value energies within \SI{-30}{\decibel} of the largest singular-value energy, providing a thresholded estimate of the number of actuator-addressable output directions retained within the domain. The condition number was computed as $\kappa=\sigma_1/\sigma_8$ using all eight singular values (Fig.~\ref{fig:plate-controllability}C).

Single virtual pixels were generated by specifying an objective $\hat{v}(\mathbf{x}_0,t)$ at location $\mathbf{x}_0$ to be a bandlimited impulse at time $t=T$ and solving for the actuator driving signals using Eq.~\ref{eq:wavefield-control}. This produced waves that converged onto location $\mathbf{x}_0$ at time $t=T$ and rapidly decayed (Fig.~\ref{fig:focusing-singlepoint}A, B). The virtual pixel area, $A$, was computed as the area enclosed by the half-maximum contour surrounding the pixel location $\mathbf{x}_0$. This procedure was repeated for each of 1225 measured locations on the display to produce a map of the virtual pixel area across the entire display (Fig.~\ref{fig:focusing-singlepoint}C). Other focusing algorithms were also investigated, such as time-reversal focusing~\cite{fink1992time} (Fig.~S6). To assess the resolvability of temporal impulses, objectives $\hat{v}(\mathbf{x}_0,t)$ were set to contain two bandlimited impulses at times $t=T$ and $t=T + \Delta t$ (Fig.~\ref{fig:focusing-singlepoint}D). $\Delta t$ was varied and temporal resolvability was assessed using a 50\% peak-to-valley criterion applied to the intensity profile, $I(t)=v(t)^2$. Two pulses were considered resolvable when the valley intensity between the peaks fell below half of the peak height. To compute multi-focal and multi-control wave field responses, the target velocity signals $\hat{v}(\mathbf{x}_c,t)$ at the $C$ pixel locations were specified to be a wide range of signals, such as impulses (Fig.~\ref{fig:focusing-kitchensink}A), noise (Fig.~\ref{fig:focusing-kitchensink}B), null (i.e., vibration suppression) (Fig.~\ref{fig:focusing-kitchensink}B), sinusoids (Fig.~\ref{fig:focusing-kitchensink}C), and temporally-sequenced pulses (Fig.~\ref{fig:focusing-kitchensink}D, E).

\subsection*{Haptic perception}
In a series of four human behavioral experiments, we evaluated how participants perceived multi-pixel tactile patterns generated on the metamaterial display using spatiotemporal inverse filtering. Participants either placed the five fingertips of their right hand on five predetermined virtual pixel locations (experiments 1, 3, and 4) or explored the five virtual pixels with their right index finger (experiment 2). The five virtual pixel locations (Fig.~\ref{fig:focusing-perception}; interdigit spacing: \SI{3.2}{\centi\meter}) were selected to accommodate a wide range of hand sizes in a natural, semi-relaxed posture and were marked with colored adhesive to aid consistent placement among participants. At the pixel locations, peak oscillation velocities across the experimental stimuli ranged from \num{1.2} to \SI{7.5}{\milli\meter/\second} (peak displacements: \num{2.2} to \SI{5.6}{\micro\meter}). Although the display produced little audible noise during operation, participants wore circumaural headphones that played pink noise for the duration of the experiments to mask any potential auditory cues. The protocol was approved by the human subjects review board at the authors' institution (Protocol No. 13-22-0139). All 10 participants (5 male, 5 female) gave their written informed consent. One participant did not participate in the final experiment due to time constraints. Detailed results and protocol information for each experiment are provided in Figs.~S9--S12 and the Supplementary Materials.

\clearpage

\newpage
\normalsize
\section*{Acknowledgments}

\noindent \textbf{Funding:} No funding was received. 
\newline

\noindent \textbf{Author Contributions:} G.R. and Y.V. designed research; G.R., M.L., and D.G. performed research; G.R., M.L., D.G., and N.T. analyzed data; G.R., N.T., and Y.V. wrote the paper.
\newline

\noindent \textbf{Competing Interests:} The authors declare that they have no competing interests.
\newline

\noindent \textbf{Data and Materials Availability:} All data needed to evaluate the conclusions in the paper are present in the main text or the supplementary materials.

\section*{Supplementary Materials}
Supplementary Methods\\
Figs. S1 to S15\\
References~\cite{cummer2016controlling, ma2016acoustic, lemoult2016soda, beli2018wave, miranda2020wave, colombi2017elastic, colquitt2017seismic, jin2021physics, rupin2015symmetry}

\newpage
\small

\clearpage
\setcounter{page}{1}
\renewcommand{\thepage}{S\arabic{page}}

\title{\bfseries\boldmath Supplementary Materials for \\ \scititle}
\maketitle

\noindent
\subsection*{This file includes:}
Supplementary Methods\\
Figs. S1 to S15\\
References~\cite{cummer2016controlling, ma2016acoustic, lemoult2016soda, beli2018wave, miranda2020wave, colombi2017elastic, colquitt2017seismic, jin2021physics, rupin2015symmetry}

\clearpage

\section*{SUPPLEMENTARY METHODS}

\subsection*{Dispersion engineering in metamaterials}
Metamaterials are artificial composite materials---typically consisting of a homogeneous host material and embedded inclusions or substructures---that exhibit unusual wave properties~\cite{cummer2016controlling, ma2016acoustic}. The two most common mechanisms for dispersion engineering in metamaterials are Bragg scattering and local resonances. Bragg scattering occurs when the operating wavelength is comparable to the lattice periodicity, making it less suitable for manipulating low-frequency, large-wavelength oscillations. In contrast, locally resonant metamaterials are governed primarily by resonator dynamics, allowing resonant inclusions, even in nonperiodic arrangements, to modify waves with wavelengths much larger than the inter-resonator spacing. Near the resonant frequency of the resonators $\omega_r$, which can be tuned independently from the host material, hybridization between the host-material dispersion branch and the local resonance produces avoided crossings, yielding upper and lower hybridized dispersion branches~\cite{lemoult2016soda}. 

The control of mechanical waves in elastic solids using locally resonant elements is an active area of metamaterials research, with major applications in low-frequency vibration attenuation~\cite{peng2015acoustic, beli2018wave, miranda2020wave} and energy localization or harvesting~\cite{chen2014metamaterials, li2017design, lee2022piezoelectric}. Rod resonators have been widely investigated for elastic metamaterials because their multiple resonances (e.g., flexural, extensional, and torsional) provide rich opportunities for tailoring plate- and surface-wave dispersion~\cite{rupin2014experimental, williams2015theory, colombi2017elastic, colquitt2017seismic, jin2021physics}. In metamaterials with multi-DOF resonators and host materials supporting multiple wave modes, local-resonator-induced hybridization can produce slow-wave branches without requiring complete band gaps~\cite{colombi2017elastic, colquitt2017seismic, rupin2015symmetry}, a behavior exhibited by the system reported here.

\subsection*{Metamaterial design: Numerical experiments}

The metamaterial design was guided by the targeted dispersion characteristics together with practical fabrication and structural constraints. Initial numerical investigations evaluated different resonator configurations and host media across several candidate materials and dimensions. A thin acrylic plate was selected as the host medium because it could be readily laser-cut to accommodate press-fit resonators. Cylindrical brass rods were ultimately selected for their favorable dispersion characteristics, relatively high density, and ease of fabrication from commercially available ultra-machinable 360 brass stock. The resonator dimensions and lattice spacing were subsequently refined together to achieve the desired dispersion characteristics while accounting for fabrication constraints. For the selected \SI{3.175}{\milli\metre}-diameter rods, the \SI{8}{\milli\metre} lattice spacing provided sufficient acrylic between adjacent press-fit holes while avoiding the increased fabrication and assembly burden associated with a denser array of smaller-diameter rods. Resonator length was also constrained to avoid excessive static deflection of the thin acrylic plate under the weight of the attached rods.

The final unit cell ($a$ = \SI{8}{\milli\meter}) consisted of a thin acrylic plate ($h_{\mathrm{plate}}$ = \SI{1.5}{\milli\meter}, $E_{\mathrm{plate}}$ = \SI{3.2}{\giga\pascal}, $\rho_{\mathrm{plate}}$ = \SI{1180}{\kilo\gram/\meter\cubed}, $\nu_{\mathrm{plate}}$ = \num{0.37}) and a cylindrical brass rod ($h_{\mathrm{rod}}$ = \SI{33.4563}{\milli\meter}, $\diameter_{\mathrm{rod}}$ = \SI{3.175}{\milli\meter}, $E_{\mathrm{rod}}$ = \SI{97}{\giga\pascal}, $\rho_{\mathrm{rod}}$ = \SI{8489}{\kilo\gram/\meter\cubed}, $\nu_{\mathrm{rod}}$ = \num{0.31}). The material properties and geometric parameters used in the numerical experiments were a combination of nominal material values and empirical measurements from fabricated components (i.e., $\rho_{\mathrm{rod}}$ and $h_{\mathrm{rod}}$ were measured empirically, with rod-height measurements shown in Fig.~S14). The acrylic plate and brass rod were both treated as linear elastic materials. Periodic boundary conditions (Floquet periodicity) were applied to the lateral faces of the acrylic plate to simulate an infinite, periodic system. All other faces were left free to oscillate. Band diagrams were computed by performing a parametric sweep along the boundary of the irreducible Brillouin zone for a square lattice, following the path $\Gamma$--$\mathrm{X}$--$\mathrm{M}$--$\Gamma$ connecting the three high-symmetry points (90 points sampled along the path; Fig.~2A). For each sampled point, the corresponding wave vector, $\mathbf{k}=[k_x,k_y]$, was assigned through the Floquet boundary conditions, and the eigenfrequencies and eigenmodes corresponding to the first $N_m$ modes were computed (Fig.~2B). The eigenfrequencies and associated wave vectors were converted into phase velocity and wavelength using the following identities: $c(f) = 2\pi f / |\mathbf{k}|$ and $\lambda(f) = c(f) / f = 2\pi/|\mathbf{k}|$, where $|.|$ denotes magnitude (Fig.~2C, D). Separately, the first resonator-associated mode frequency was computed via unit-cell eigenfrequency analysis with fixed rather than periodic boundary conditions. This yielded an effective resonance frequency of the rod--plate unit-cell structure, $f_n=\omega_r/2\pi=\SI{781}{\hertz}$, consistent with the behavior observed in Fig.~2B.

\subsection*{Mass loading and mode hybridization in slow-wave dispersion}

Attaching the brass rod resonators to the acrylic plate introduces additional mass and modifies wave dispersion through coupling between plate and resonator modes. To assess the contribution of mass loading to wavelength reduction, we compared the numerical dispersion of the metamaterial plate with that of a hypothetical, not physically realizable, homogeneous plate having the same areal mass density as the metamaterial but retaining the bending rigidity of the original acrylic plate.

The total mass of the metamaterial unit cell was calculated from the geometry and densities of its constituent materials. The brass rod has a mass of $m_{\mathrm{rod}} = \rho_{\mathrm{rod}}\pi d_{\mathrm{rod}}^2 h_{\mathrm{rod}}/4 \approx$ \SI{2.2486}{\gram}, while the acrylic plate section, accounting for the hole occupied by the rod, has a mass of $m_{\mathrm{plate}} = \rho_{\mathrm{plate}}(a^2-\pi d_{\mathrm{rod}}^2/4)h_{\mathrm{plate}} \approx$ \SI{0.0993}{\gram}. The total metamaterial unit-cell mass is therefore $m_{\mathrm{meta}} = m_{\mathrm{plate}}+m_{\mathrm{rod}} \approx$ \SI{2.35}{\gram}. To construct a hypothetical equivalent-mass plate, we distribute the total metamaterial unit-cell mass uniformly over the original plate thickness, giving an equivalent density of
\begin{equation}
\rho_\mathrm{equiv} = \frac{m_\mathrm{meta}}{a^2h_\mathrm{plate}} \approx \SI{24457}{\kilo\gram/\meter\cubed}.
\end{equation}
Substituting this density into Eq.~4, while retaining the original acrylic plate's thickness and elastic properties, yields the phase velocity of the hypothetical equivalent-mass plate. Compared with the original acrylic density, $\rho_{\mathrm{plate}}$ = \SI{1180}{\kilogram\per\cubic\metre}, the equivalent density is approximately 20.7 times greater. Because wavelength scales with density as $\lambda \propto \rho^{-1/4}$ at fixed frequency and bending rigidity, this mass increase predicts a $20.7^{1/4} \approx 2.13$-fold reduction in wavelength relative to the homogeneous acrylic plate.

We compared the resulting dispersion relation of the hypothetical equivalent-mass plate with that of the metamaterial plate (Fig.~S1). At low frequencies ($<$ \SI{75}{\hertz}), the dispersion behavior of the two plates is largely similar, but the effects of mode hybridization are already visible at \SI{75}{\hertz} and become increasingly pronounced with increasing frequency. At \SI{300}{\hertz}, the predicted wavelength of the equivalent-mass plate is approximately \SI{5.9}{\centi\metre}, compared with \SI{3.0}{\centi\metre} for the metamaterial and \SI{12.7}{\centi\metre} for the homogeneous acrylic plate. This additional wavelength reduction is associated with the downward shift of the fundamental dispersion branch arising from coupling between plate and resonator modes (Fig.~2B).

The hypothetical equivalent-mass plate is intended to isolate the effects of mass loading by assuming that the areal mass density can be increased without altering the bending rigidity of the original acrylic plate. However, this combination of density and elastic properties does not correspond to a physically realizable material. For comparison, we additionally considered a homogeneous gold plate with the same areal mass density as the metamaterial. We selected gold owing to its high mass density and relatively low elastic modulus compared with other materials of similar density (e.g., tungsten). We used nominal gold properties of $\rho_{\mathrm{gold}}$ = \SI{19300}{\kilogram\per\cubic\metre}, $E_{\mathrm{gold}}$ = \SI{78}{\giga\pascal}, and $\nu_{\mathrm{gold}}$ = 0.44. Matching the areal mass density of the metamaterial requires a gold plate thickness of $h_{\mathrm{gold}}=\rho_{\mathrm{equiv}}h_{\mathrm{plate}}/\rho_{\mathrm{gold}} \approx$ \SI{1.9}{\milli\metre}. We calculated the corresponding dispersion relation using Eq.~4 (Fig.~S1). The equivalent-mass gold plate exhibits longer wavelengths than even the original homogeneous acrylic plate. This result highlights the limitations of treating mass loading as an independent physical mechanism, as changes in mass, bending rigidity, and modal behavior are coupled in realizable structures.

\subsection*{Device fabrication}
To construct the rod resonators, brass rods were cut to the desired height from stock material (Ultra-Machinable 360 Brass Rod, McMaster-Carr, USA) using a manual rod cutter, then sanded and polished. The finished rods were measured using calipers (average height: \SI{33.46}{\milli\meter}, diameter: \SI{3.175}{\milli\meter}; Fig.~S14). A square lattice ($a$ = \SI{8}{\milli\meter}) of holes (diameter: \SI{3.09}{\milli\meter}) was laser cut through the acrylic plate, and the brass rods were press-fit into the holes after applying a thin layer of epoxy around the circumference of each rod.

The metamaterial and homogeneous plates were supported by a 3D-printed assembly consisting of eight support pillars with square cross-sections (\num{1.8} $\times$ \SI{1.8}{\centi\meter}) and eight voice-coil mounting stands. Polyurethane foam blocks (\num{1.8} $\times$ \num{1.8} $\times$ \SI{1}{\centi\meter}) were placed atop each pillar, and the plates rested on these blocks. To fabricate each voice coil, 30-AWG magnet wire was wrapped around a thin-shell aluminum bobbin (950 turns) and attached to a ceramic heat sink using thermally conductive epoxy (8329TFF, MG Chemicals, Canada) to improve heat dissipation. Two NdFeB magnets (diameter: \SI{1}{\centi\meter}, thickness: \SI{0.5}{\centi\meter}) were placed on either side of the plate, held together by magnetic attraction, and positioned directly above the voice coil. 

The response of each actuator was measured using an LDV focused on the NdFeB magnet positioned on the plate's top surface. The frequency response was measured using a \SI{3}{\second} linear sine sweep from \num{50} to \SI{625}{\hertz}. A Kirkeby inverse filter with trapezoidal regularization was computed from this response and applied to each actuator to flatten its driven response (Fig.~S3). All impulse-response measurements and experiments were performed with actuator compensation applied during driving.

\subsection*{Additional experimental characterization}

\subsubsection*{Experimental dispersion spectra}
Experimental dispersion spectra were computed from the measured Green's functions $g_i(\mathbf{x},t)$ for each of the eight actuators. For each actuator, the Green's functions were first Fourier transformed in time to obtain complex-valued spatial wave fields at each frequency. These wave fields were multiplied by a two-dimensional Tukey window (taper fraction 0.3) and subsequently Fourier transformed in space to obtain the corresponding wavenumber spectra. The magnitude spectra were radially averaged in wavenumber space, using $k = |\mathbf{k}|=\sqrt{k_x^2+k_y^2}$, to incorporate contributions from different propagation directions rather than restricting the analysis to either the $k_x$ or $k_y$ axis. The resulting spectra were averaged across the eight actuators and normalized to their maximum magnitude to obtain the experimental dispersion spectra for both plates (Fig.~S4). Unlike the wavelength-estimation procedure described in Materials and Methods, these spectra were computed without Laplacian-of-Gaussian spatial filtering. The overlaid wavenumber estimates correspond to the wavelength estimates reported in Fig.~\ref{fig:plate-comparison}A, converted using $|\mathbf{k}|=2\pi/\lambda$.

For both the metamaterial and homogeneous plates, the wavenumber estimates closely follow the dominant features in the experimental dispersion spectra within the \num{75}--\SI{400}{\hertz} operating band (Fig.~S4). For the metamaterial plate, the spectral response decreases at higher frequencies, making the dominant dispersion branch increasingly difficult to identify. Above approximately \SI{400}{\hertz}, the branch can no longer be tracked with confidence. These results are consistent with the reduction in driven response (Fig.~\ref{fig:plate-comparison}B) and the observed deviation of the wavelength estimates from their numerical predictions beginning at around \SI{400}{\hertz} (Fig.~\ref{fig:plate-comparison}A).

\subsubsection*{Attenuation time constant}
The attenuation time constant characterizes how long vibrations persist following excitation and is an important parameter in wave-field control methods such as inverse filtering. In particular, the duration of the impulse response influences the required inverse-filter length and can affect the latency of real-time implementations~\cite{hudin2026spatio}. We therefore characterized the vibration attenuation of the metamaterial and homogeneous plates using their measured Green's functions.

To estimate these time constants, the root-mean-square (RMS) response was computed across all measured surface locations and actuators. An exponential decay model, $A(t)=A_0e^{-t/\tau}$, was fit to the resulting RMS response using ordinary least squares in logarithmic space over the interval $0 \leq t \leq T_\mathrm{fit}$~\cite{hudin2026spatio}. To evaluate the sensitivity of the estimated time constant $\tau$ to the fitting interval, the analysis was repeated with fitting-window durations $T_\mathrm{fit}$ ranging from \num{5} to \SI{200}{\milli\second}. The estimated time constants were relatively stable over fitting-window durations of \num{20}--\SI{100}{\milli\second} for the metamaterial plate and \num{60}--\SI{100}{\milli\second} for the homogeneous plate (Fig.~S15A). For longer fitting windows, the estimated time constants increased, potentially reflecting deviations from single-exponential decay and the growing influence of measurement noise at later times. Given this fitting-window dependence, we used the median estimates within the respective stable ranges to characterize each plate. The metamaterial plate exhibited a moderately longer attenuation time constant of approximately \SI{30.8}{\milli\second}, compared with \SI{22.0}{\milli\second} for the homogeneous plate (Fig.~S15B). Thus, for the architecture investigated here, introducing the resonator array improved spatial resolution but also increased vibration persistence, potentially imposing additional constraints on the latency of real-time wave-field control.

\subsection*{Data-driven wave-field control: Regularization}

To compute our actuator driving signals (Eq.~8), we employed frequency-dependent Tikhonov regularization to limit amplification of weakly responding spatial components. The regularization parameter $\alpha(\omega)$ followed a trapezoidal profile, with $\alpha$ = \num{1} over \num{70}--\SI{450}{\hertz} and \SI{15}{\hertz} transition regions spanning \num{55}--\SI{70}{\hertz} and \num{450}--\SI{465}{\hertz}. Stronger regularization ($\alpha=10^6$) was applied below \SI{55}{\hertz} and above \SI{465}{\hertz} to suppress unwanted spectral contributions. The same regularization profile was employed across all experiments, with the frequency content of individual stimuli determined separately through their prescribed waveforms and subsequent zero-phase bandpass filtering.

Within the operating band, the nonzero eigenvalues of $\mathbf{G}_C^\dagger\mathbf{G}_C$ were several orders of magnitude greater than $\alpha^2$ for most control objectives, so regularization contributed little to the resulting solution. However, for certain control configurations at lower frequencies and near the upper edge of the operating bandwidth (approximately \SI{400}{\hertz}), the smallest nonzero eigenvalues were comparable in magnitude to $\alpha^2$, so regularization limited the actuator driving amplitudes associated with these weakly responding directions. The regularization parameter was selected heuristically, and its numerical value depends on the scaling of the measured Green's functions. Accordingly, the parameter should be adjusted when applying this procedure to differently scaled propagation matrices.

\subsection*{Haptic perception}
In each experiment, participants were asked to respond verbally, and an experimenter entered their responses on a computer graphical user interface. The experimenter was always blind to the correct response. To facilitate familiarization with the sensations they would feel in each experiment, participants felt all stimuli in a given experiment (in a randomized order) before the experimental trials began. However, participants were unaware that these first trials were part of the familiarization phase and were not informed whether their answers in this portion of the experiment were correct. These trials were removed during analysis. Following the familiarization phase, all experimental trials were completely randomized.

In all experiments, spatiotemporal inverse filtering was used to compute actuator driving signals to generate tactile feedback at the five virtual pixel locations. For each stimulus set in a given experiment, actuator driving signals were normalized to a prescribed peak voltage using a common scaling factor. The selected voltages varied between experiments because some spatial and temporal stimulus patterns required greater actuator effort than others, and because the experiments employed different types of stimuli. Amplitudes were selected to provide readily perceptible tactile feedback that could be experienced comfortably for the duration of the experiments.

In the first behavioral experiment ($n=\num{10}$; Fig.~S9), participants placed the five fingertips of their right hand on five predetermined virtual pixel locations. On each trial, one virtual pixel delivered a train of short white-noise bursts (\SI{25}{\milli\second}, \SI{75}{\milli\second} pause, 5 bursts), while the other four were actively suppressed (i.e., assigned null targets). Participants identified the active pixel. The experiment included five stimulus conditions, corresponding to the five possible active-pixel locations (Fig.~S9B). Participants were free to repeat each stimulus as many times as they preferred before providing their verbal response. Each condition was repeated 10 times, yielding 50 experimental trials in addition to 5 familiarization trials. 

In the second behavioral experiment ($n=\num{10}$; Fig.~S10), participants used the index finger of their right hand to explore five virtual pixels and identify the pixel that presented a distinct feedback pattern (``odd-one-out'' paradigm). On each trial, participants were required to feel each of the five pixels at least once, but their search technique, dwell time at each pixel, and decision to revisit previous pixels were otherwise unconstrained. For this experiment, we also recorded the time required to complete each trial for 8 of the 10 participants (Fig.~S10C). The experiment included 10 stimulus conditions: five possible odd-one-out locations $\times$ two choices for which feedback pattern served as the odd-one-out (Fig.~S10B). The two feedback patterns were a \SI{200}{\hertz} sinusoidal wavelet train (pattern 1; \SI{20}{\milli\second} duration, \SI{80}{\milli\second} pause, five wavelets) and a \SI{90}{\hertz} sinusoid (pattern 2; \SI{425}{\milli\second} duration). Because the two feedback patterns differed in frequency content, their relative amplitudes were adjusted to ensure that both were readily perceptible without either dominating the other. After a brief pause, these patterns continued to repeat until participants responded verbally. Each condition was repeated five times, yielding 50 experimental trials in addition to 10 familiarization trials.

In the third behavioral experiment ($n=\num{10}$; Fig.~S11), participants once again placed the five fingertips of their right hand on five predetermined virtual pixel locations. Motion cues were delivered by sequentially activating the pixels beneath each finger. Participants identified the direction of sequential pixel activation as either rightward (thumb to pinky) or leftward (pinky to thumb). Feedback consisted of non-overlapping white-noise bursts, each having a duration equal to the interval $\Delta t$ between successive pixel activations. This interval ranged from \SI{640}{\milli\second} to \SI{80}{\milli\second}, corresponding to motion cue speeds ranging from \SI{5}{\centi\meter/\second} to \SI{40}{\centi\meter/\second}. The actuator driving signals were normalized separately for each speed, resulting in modest variations in peak vibration velocity across speeds. However, the same scaling factor was applied to both directions at each speed, ensuring consistent amplitudes and preventing amplitude from providing a systematic cue to motion direction. There were 8 distinct stimuli in the experiment (Fig.~S11B)---4 different motion speeds and 2 directions per speed---and each stimulus was repeated 10 times, yielding 80 experimental trials in addition to 8 familiarization trials. Within each trial, the motion sequence was replayed 1, 2, 3, or 4 times, depending on speed, with a \SI{500}{\milli\second} pause between repetitions, to approximately compensate for the shorter stimulus duration at higher speeds. Participants were free to request additional repetitions of the full trial stimulus before providing their response.

In the final behavioral experiment ($n=\num{9}$; Fig.~S12), participants used the same hand posture and setup as in experiments 1 and 3. Feedback was presented simultaneously at all five virtual pixels. Similar to experiment 2, one pixel delivered a distinct feedback pattern (\SI{200}{\hertz} sinusoidal wavelet train, \SI{20}{\milli\second} duration, \SI{80}{\milli\second} pause, five wavelets), while the other four pixels delivered a different pattern (\SI{90}{\hertz} sinusoid, \SI{425}{\milli\second} duration). The amplitude settings for both feedback patterns were identical to those in experiment 2, where they were adjusted to account for differences in frequency content. Participants identified the ``odd-one-out'' pixel. The experiment included five stimulus conditions, corresponding to the five possible locations of the distinct pattern (Fig.~S12B). Each condition was repeated 10 times, yielding 50 experimental trials in addition to 5 familiarization trials. Participants were free to request additional repetitions of the stimulus before providing their response.

\clearpage

\section*{SUPPLEMENTARY FIGURES}

\makeatletter
\setlength{\@fptop}{0pt plus 1fil}   
\setlength{\@fpsep}{12pt plus 1fil}  
\setlength{\@fpbot}{0pt plus 1fil}   
\makeatother

\vspace*{15mm}
\begin{figure*}[h]
    \centering
    \includegraphics[scale=0.9]{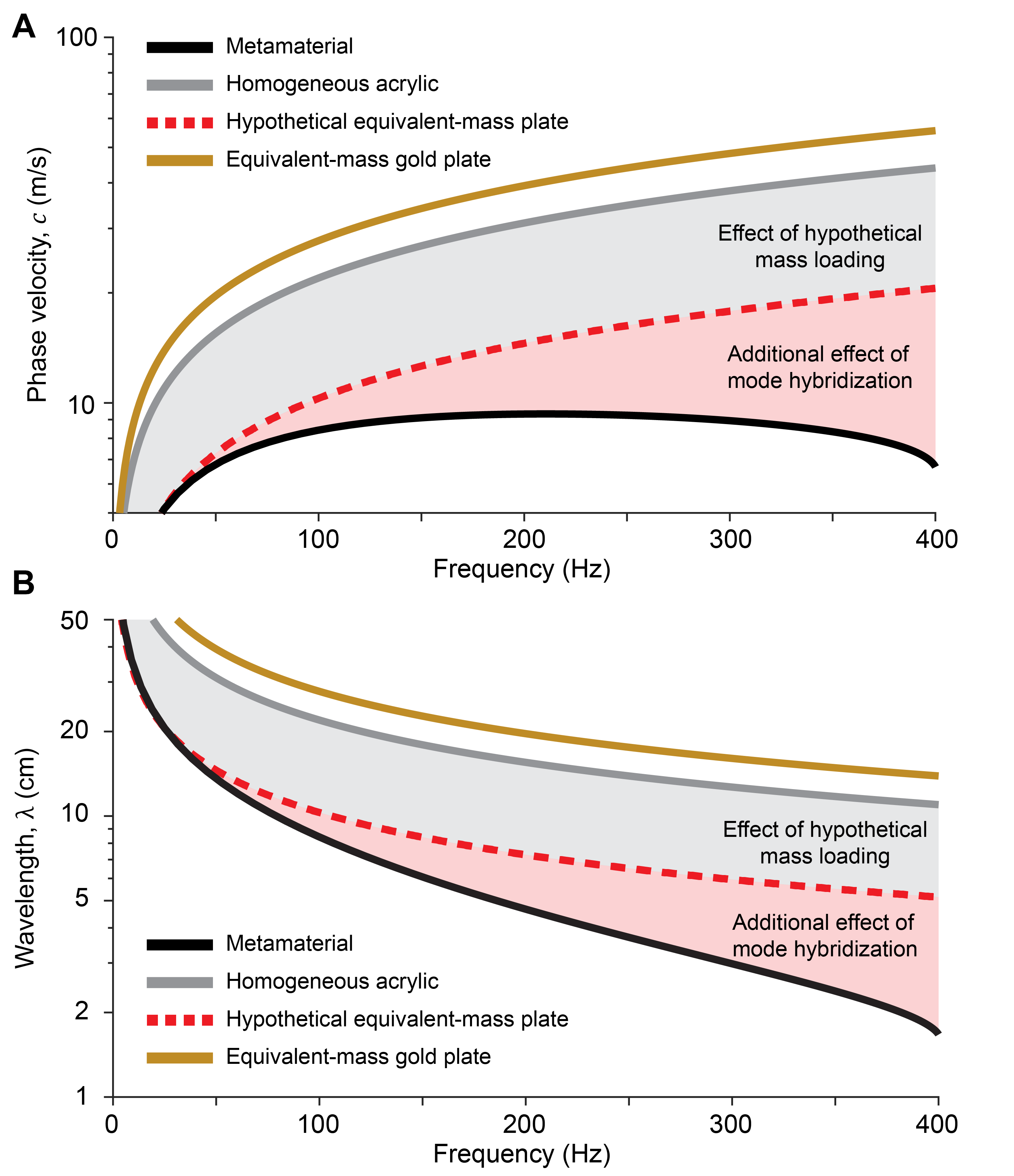}
    \caption{\textbf{Fig. S1. Mass loading and resonator-induced mode hybridization in the slow-wave dispersion of the metamaterial plate.} (\textbf{A}) Frequency-dependent phase velocity $c$ and (\textbf{B}) wavelength $\lambda$ for the metamaterial plate (black; $m = 1$ along $\Gamma-\mathrm{X}$), homogeneous acrylic plate (gray), hypothetical equivalent-mass plate (dashed red), and equivalent-mass gold plate (gold). The hypothetical equivalent-mass plate, which is not physically realizable, retains the thickness and elastic properties of the acrylic plate while matching the areal mass density of the metamaterial plate. Shaded regions indicate the reduction associated with hypothetical mass loading (gray) and the additional reduction associated with resonator-induced mode hybridization (red).}
    \label{fig:supp-hybridization}
\end{figure*}

\vspace*{15mm}
\begin{figure*}[h]
    \centering
    \includegraphics[scale=0.9]{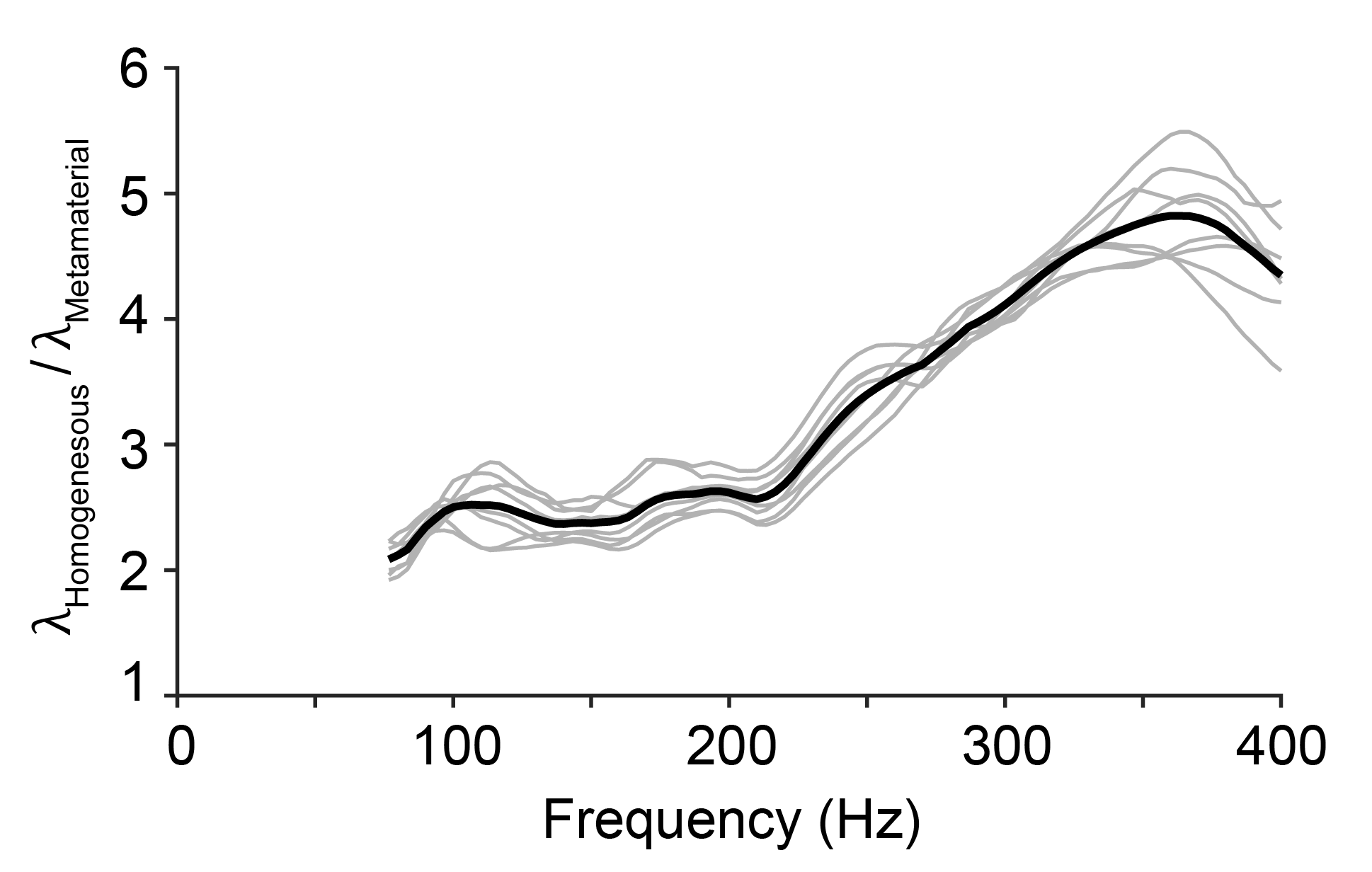}
    \caption{\textbf{Fig. S2.} Frequency-dependent wavelength reduction factor computed from the measured homogeneous acrylic and metamaterial plate wavelengths in Fig.~4A, defined as $\lambda_{i,\mathrm{homogeneous}}(f)/\lambda_{i,\mathrm{metamaterial}}(f)$, where $i$ indexes the driven actuator. Black line: mean across actuators; gray lines: individual actuator-driven wavelength ratios.}
    \label{fig:ch6-supp-wavelength}
\end{figure*}

\begin{figure*}[p]
    \centering
    \includegraphics[scale=0.9]{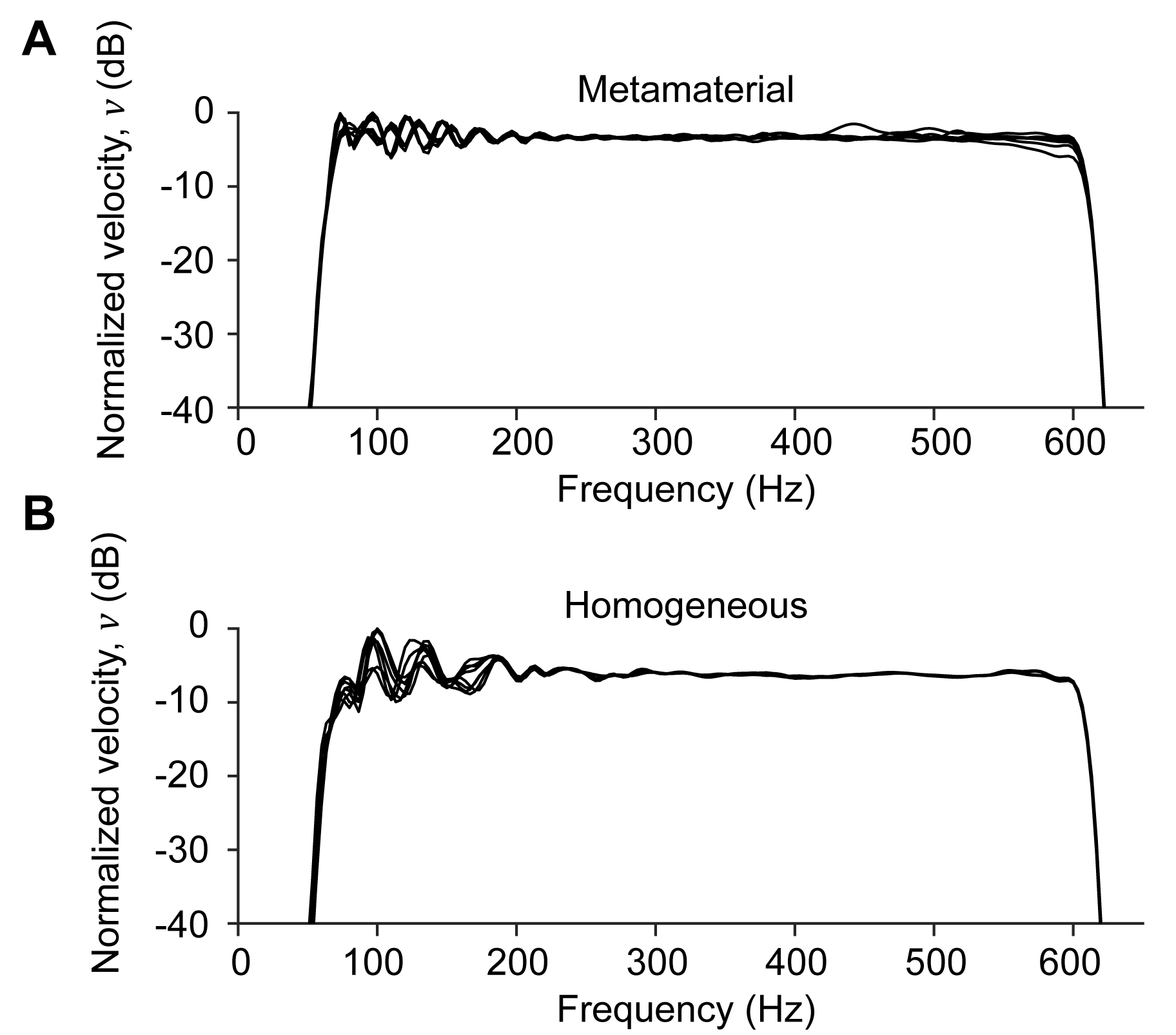}
    \caption{\textbf{Fig. S3.} Compensated actuator frequency responses measured on the NdFeB magnets for the metamaterial plate configuration (\textbf{A}) and homogeneous acrylic plate configuration (\textbf{B}). These responses were used to normalize the driven plate response shown in Fig.~4B.}
    \label{fig:ch6-supp-actuator-res}
\end{figure*}

\begin{figure*}[p]
    \centering
    \includegraphics[scale=0.9]{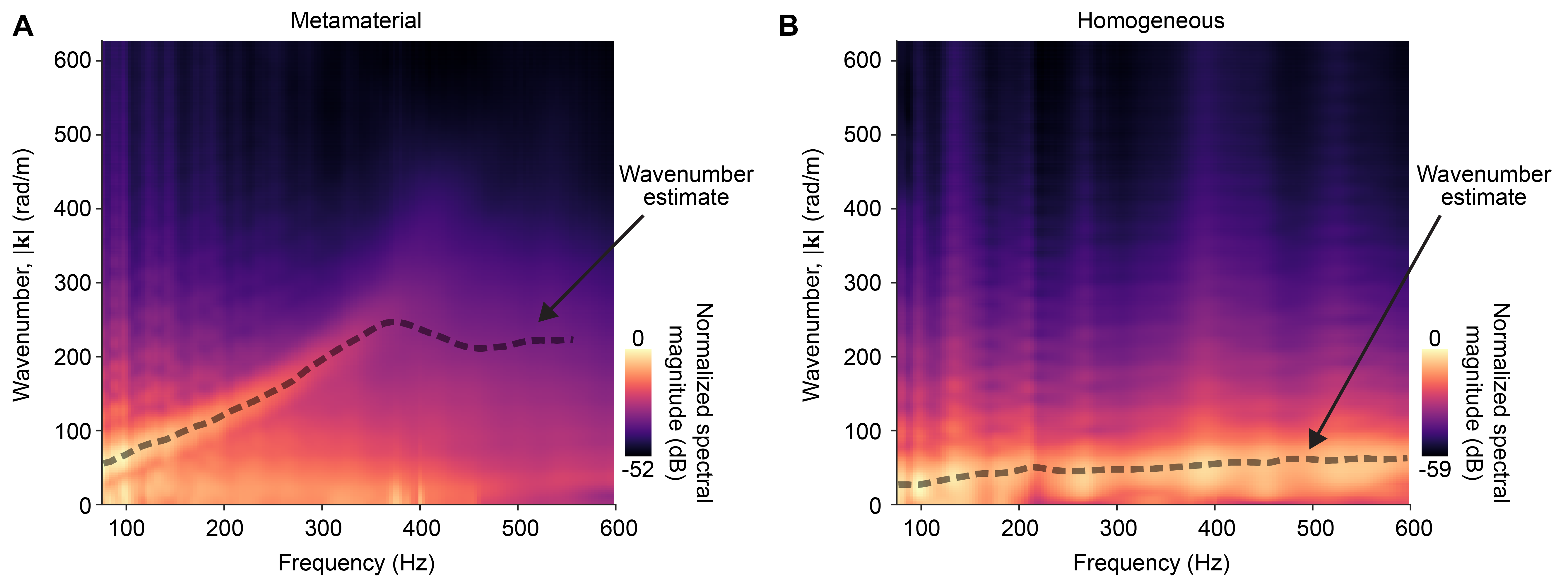}
    \caption{\textbf{Fig. S4. Experimental dispersion spectra of the metamaterial and homogeneous acrylic plates.} Frequency–wavenumber spectra computed from the measured Green's functions for (\textbf{A}) the metamaterial plate and (\textbf{B}) the homogeneous acrylic plate using spatial and temporal Fourier transforms. Black dashed lines indicate wavenumbers calculated from the experimental wavelength estimates ($|\textbf{k}| = 2\pi/\lambda$; Fig.~4A).}
    \label{fig:supp-dispersion-spectra}
\end{figure*}
\vspace*{\fill}

\begin{figure*}[p]
    \centering
    \includegraphics[scale=0.9]{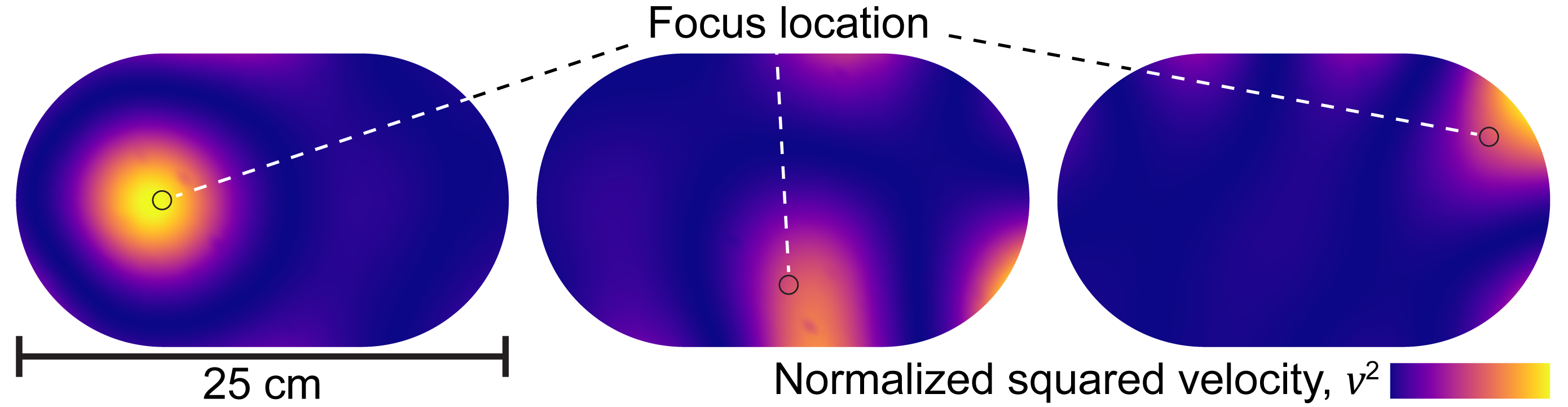}
    \caption{\textbf{Fig. S5. Extended single pixel results on the homogeneous acrylic plate.} Virtual pixels were generated at various locations on the homogeneous plate (wave field at focusing frame $t=0$ shown). At some display locations (left panel), this yielded a defined but large-area focus. At other locations, the homogeneous plate failed to produce a well-defined focus (middle and right panels).}
    \label{fig:ch6-supp-homogeneous}
\end{figure*}

\begin{figure*}[p]
    \centering
    \includegraphics[scale=0.9]{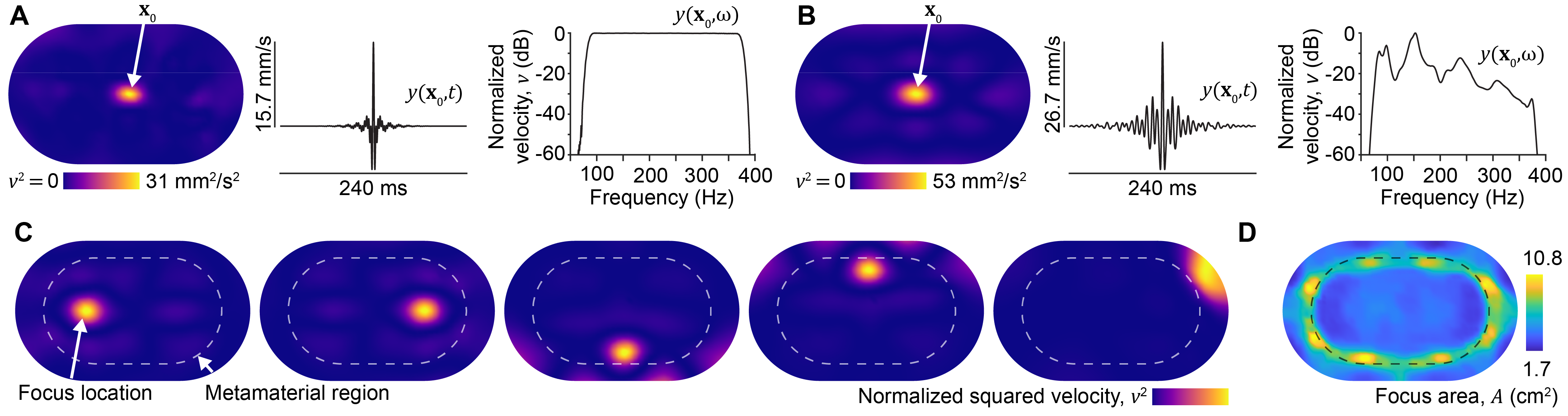}
    \caption{\textbf{Fig. S6. Extended single pixel results on the metamaterial plate.} (\textbf{A}) Spatiotemporal inverse filtering can generate virtual pixels with precisely controlled frequency content (target waveform: bandlimited impulse). Left panel: wave field at focusing frame $t=0$. Right panels: time- and frequency-domain response at the pixel location. (\textbf{B}) Time-reversal focusing produces higher peak velocities at the pixel location, but at the cost of controlled frequency content. Left panel: wave field at focusing frame $t=0$. Right panels: time- and frequency-domain response at the focus location. (\textbf{C}) Virtual pixels generated at different display locations using time-reversal focusing (focusing frame $t=0$ shown). (\textbf{D}) Spatial map of the virtual pixel area $A$ on the metamaterial plate when forming pixels at locations across the display using time-reversal focusing. Owing to plate attenuation at higher frequencies, time-reversal focusing yields virtual pixel areas that are, on average, slightly larger than those produced using spatiotemporal inverse filtering.}
    \label{fig:ch6-supp-trf}
\end{figure*}

\begin{figure*}[p]
    \centering
    \includegraphics[scale=0.9]{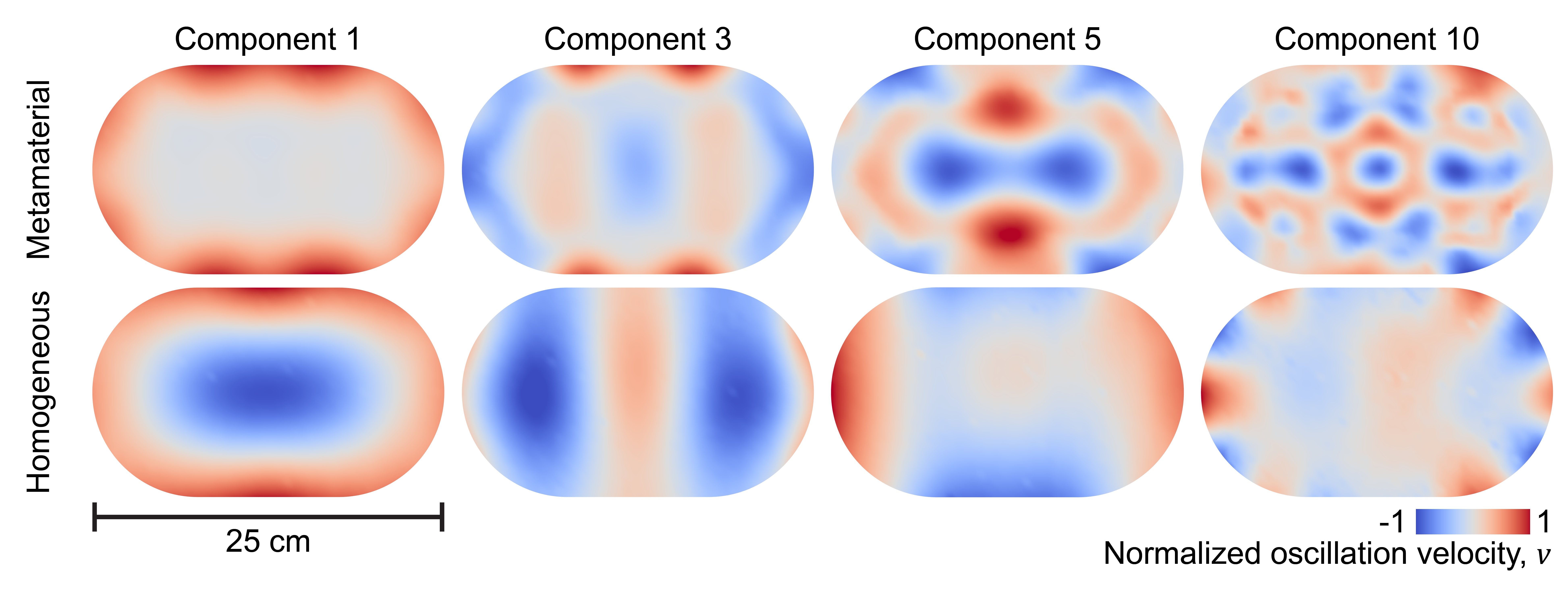}
    \caption{\textbf{Fig. S7.} Representative spatial POD components computed from the measured impulse-response matrix $\mathbf{H}$ for the metamaterial plate (top row) and homogeneous acrylic plate (bottom row). Components 1, 3, 5, and 10 are shown, illustrating the shorter-scale spatial structure in the metamaterial response.}
    \label{fig:ch6-supp-podcomponents}
\end{figure*}

\begin{figure*}[p]
    \centering
    \includegraphics[scale=0.9]{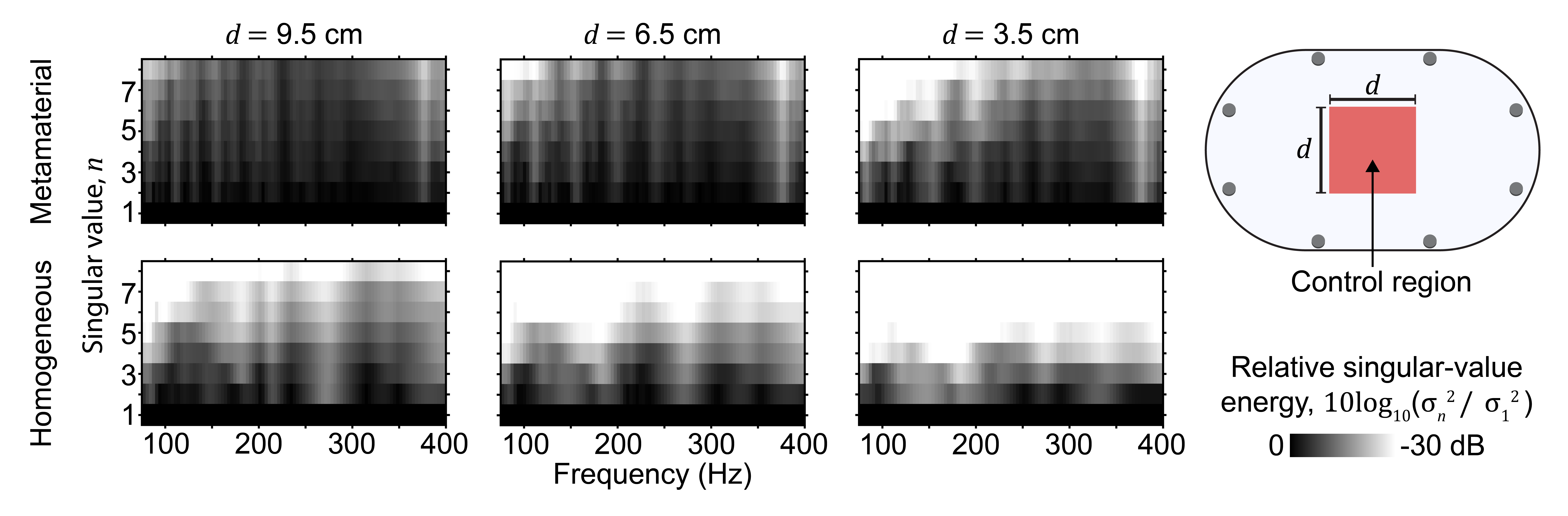}
    \caption{\textbf{Fig. S8.} Frequency-dependent relative singular-value energy of the domain-restricted propagation matrix $\mathbf{G}_D(\omega)$ for the metamaterial and homogeneous acrylic plates. Relative singular-value energy was computed as $10\log_{10}(\sigma_i^2/\sigma_1^2)$ for square domains of edge length $d=\num{9.5}$, \num{6.5}, and \SI{3.5}{\centi\meter}; these spectra were used to compute the band-averaged summaries presented in Fig.~6B.}
    \label{fig:ch6-supp-freqdep-control}
\end{figure*}

\begin{figure*}[p]
    \centering
    \includegraphics[scale=0.9]{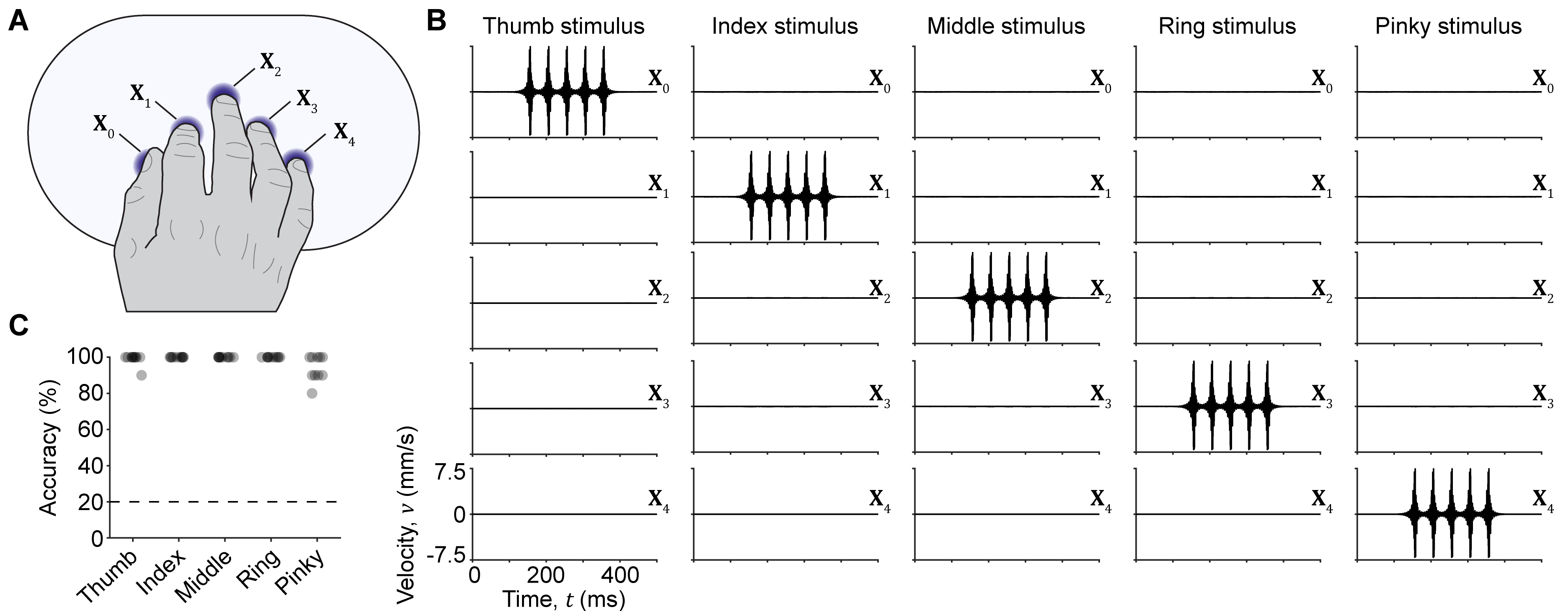}
    \caption{\textbf{Fig. S9. Extended behavioral experiment results for the pixel identification task.} (\textbf{A}) Participants placed the five fingertips of their right hand on specified virtual pixel locations and identified the active pixel. (\textbf{B}) Time-domain responses at the virtual pixel locations for the five stimuli used in the experiment. (\textbf{C}) Results ($n = 10$) of the behavioral experiment (per-subject accuracies shown; dashed line: chance accuracy).}
    \label{fig:ch6-supp-oneofN}
\end{figure*}

\begin{figure*}[p]
    \centering
    \includegraphics[scale=0.9]{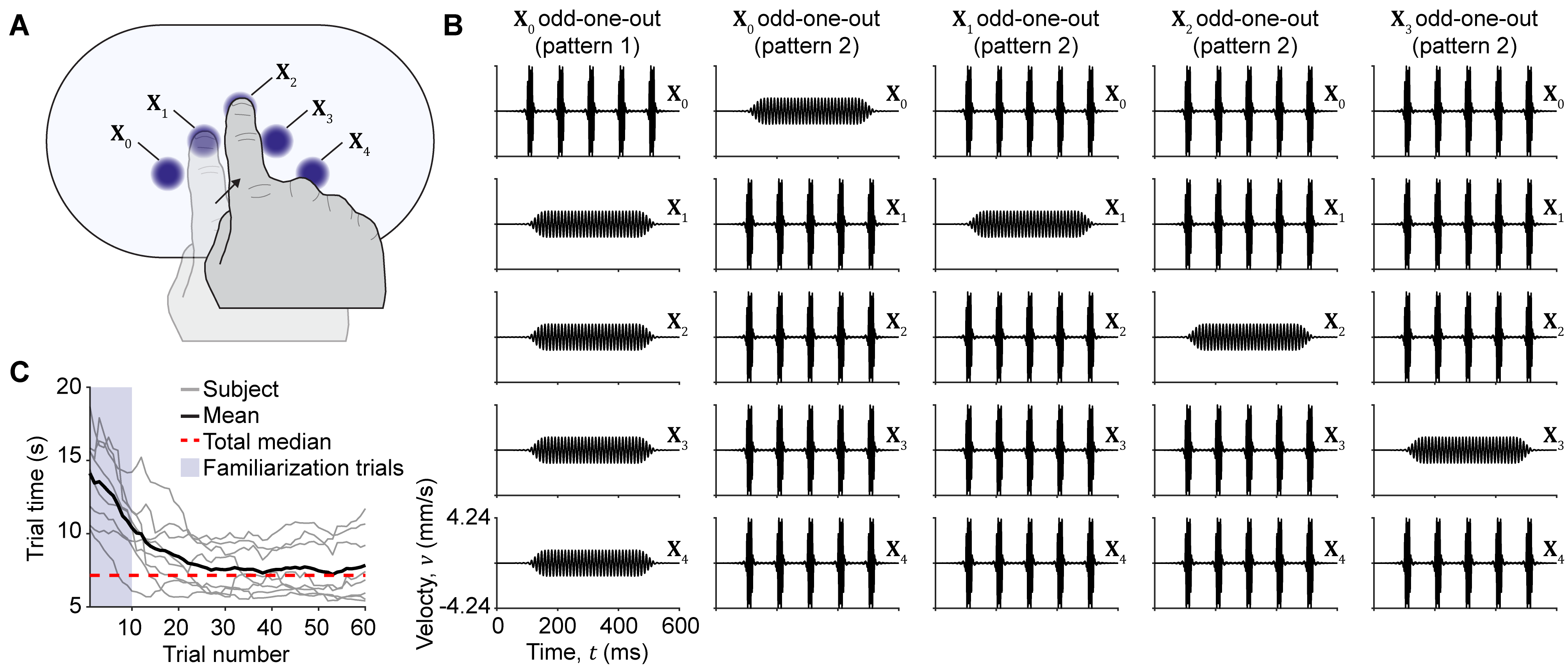}
    \caption{\textbf{Fig. S10. Extended behavioral experiment results for the haptic exploration task.} (\textbf{A}) Participants explored five virtual pixel locations and identified the one that exhibited a distinct vibration pattern. (\textbf{B}) Time-domain responses at the virtual pixel locations for a subset of stimuli used in the experiment. There were five possible odd-one-out locations and two choices for which feedback pattern served as the odd-one-out (5 $\times$ 2 = 10 distinct patterns; 5 of 10 patterns shown). (\textbf{C}) Participant accuracy was 100\% in this experiment ($n = 10$). We also captured the trial times for 8 of 10 participants. Participant trial times generally decreased as the experiment proceeded. By the end of the experiment, participants were able to rapidly search the pixels and identify the odd-one-out in under \SI{8}{\second}.}
    \label{fig:ch6-supp-oddballsearch}
\end{figure*}

\begin{figure*}[p]
    \centering
    \includegraphics[scale=0.9]{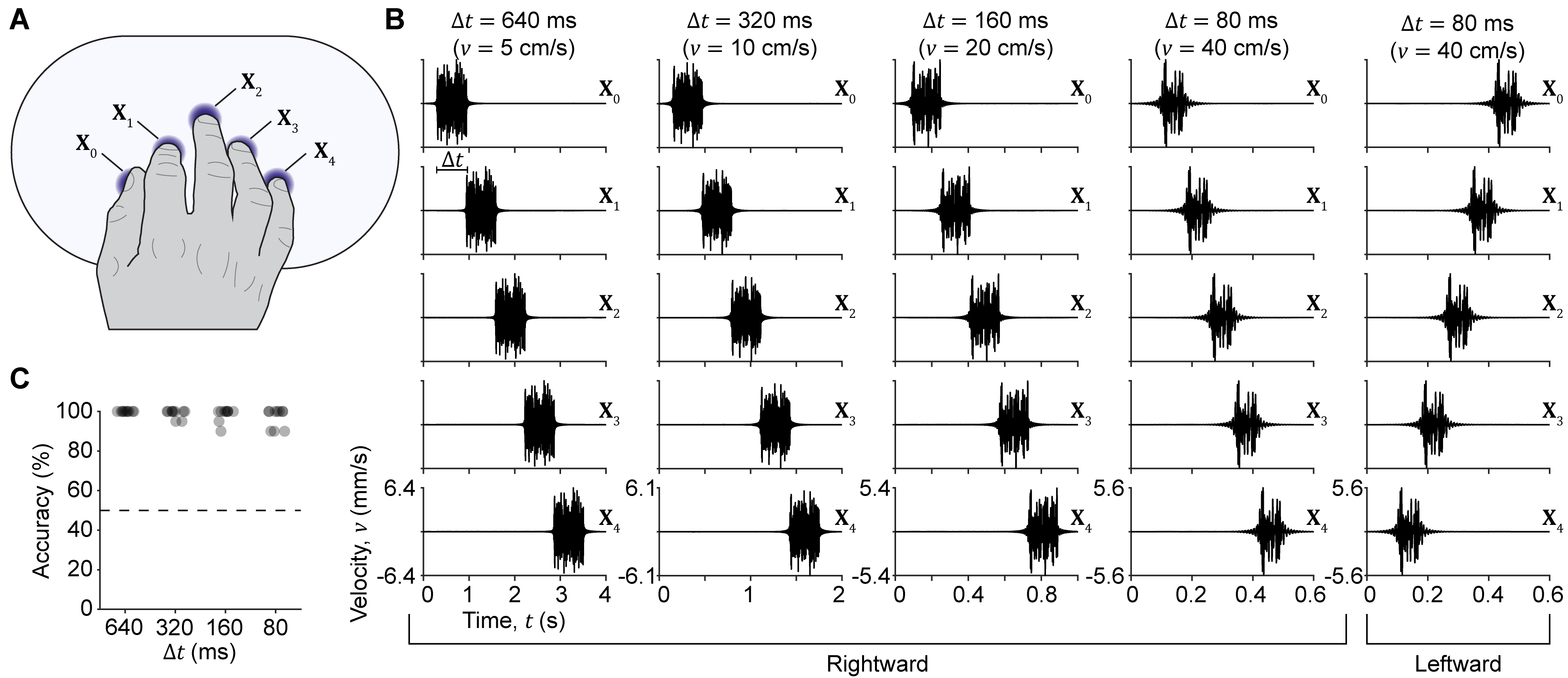}
    \caption{\textbf{Fig. S11. Extended behavioral experiment results for the motion direction identification task.} (\textbf{A}) Participants placed the five fingertips of their right hand on specified virtual pixel locations and identified the direction of sequentially activated pixels in a binary forced-choice paradigm. (\textbf{B}) Time-domain responses at the virtual pixel locations for a subset of stimuli used in the experiment. Four different motion speeds ranging from \SI{5}{\centi\meter/\second} to \SI{40}{\centi\meter/\second} were tested (4 speeds $\times$ 2 directions = 8 distinct patterns; 5 of 8 patterns shown). (\textbf{C}) Results ($n = 10$) of the behavioral experiment (per-subject accuracies shown; dashed line: chance accuracy).}
    \label{fig:ch6-supp-motion}
    \vspace{0.1\textheight}

\end{figure*}

\begin{figure*}[p]
    \centering
    \includegraphics[scale=0.9]{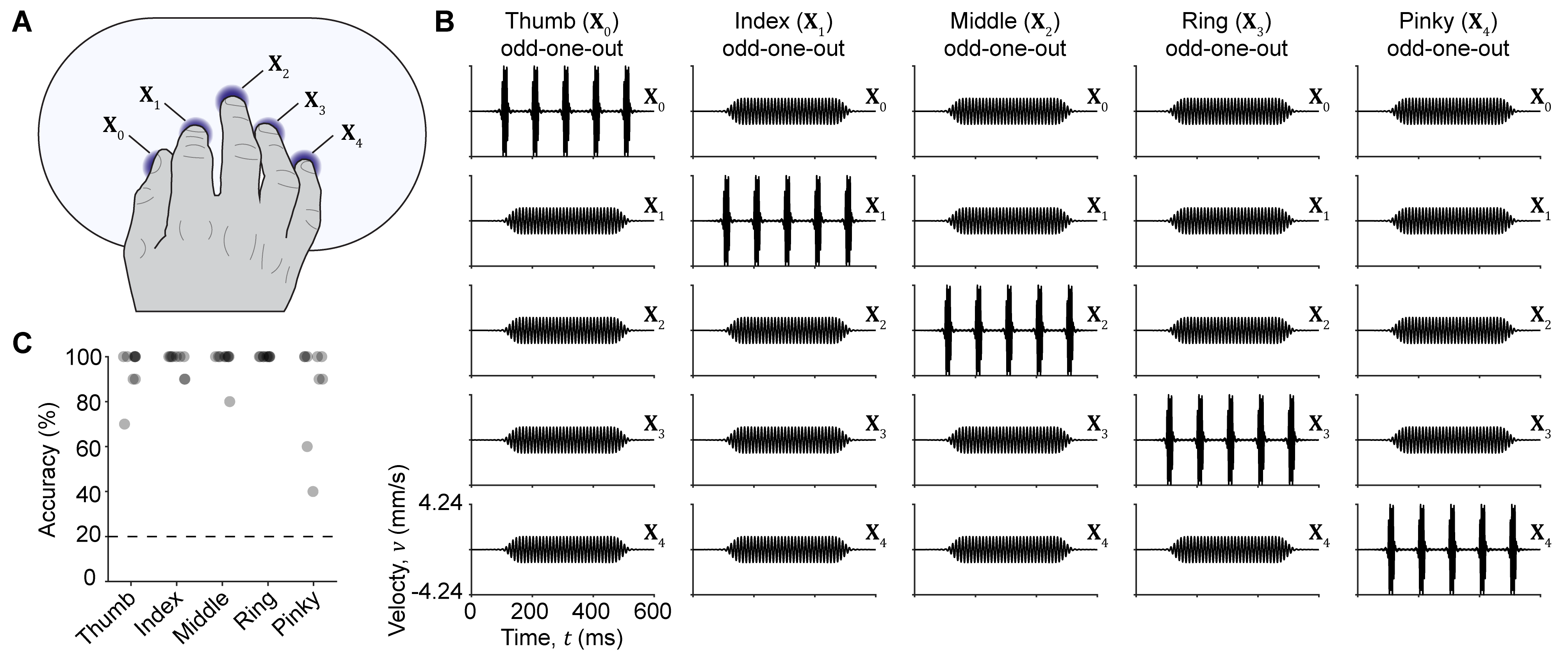}
    \caption{\textbf{Fig. S12. Extended behavioral experiment results for multi-touch odd-one-out task.} (\textbf{A}) Participants placed the five fingers of their right hand on specified virtual pixel locations and identified the location that presented a distinct vibration pattern. (\textbf{B}) Time-domain responses at the virtual pixel locations of the five stimuli employed in the experiment. (\textbf{C}) Results ($n = 9$) of the behavioral experiment (per-subject accuracies shown; dashed line: chance accuracy).}
    \label{fig:ch6-supp-oddballmultipoint}
    \vspace{0.1\textheight}
\end{figure*}

\afterpage{\clearpage}
\begin{figure*}[p]
    \centering
    \includegraphics[scale=0.9]{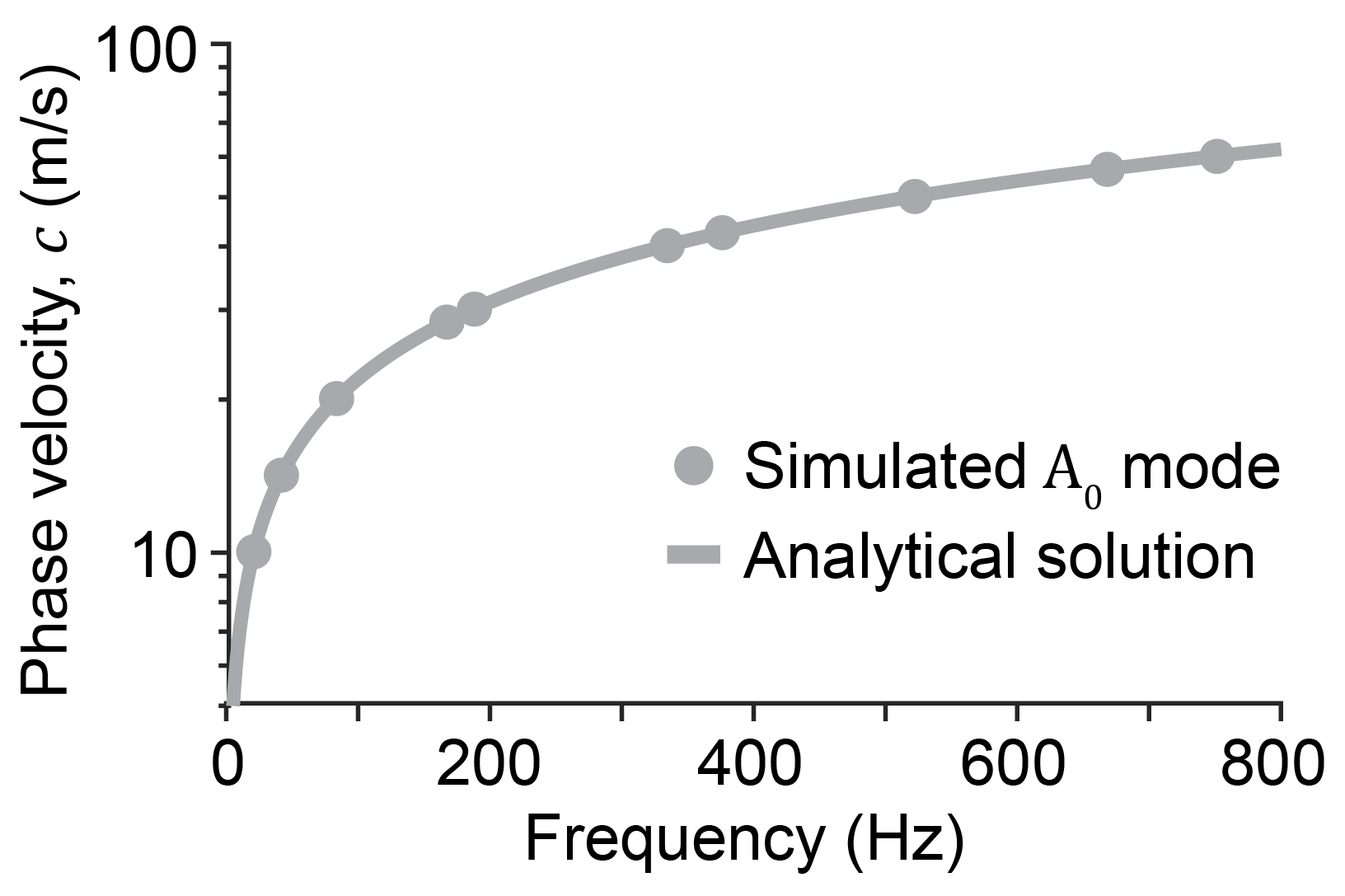}
    \caption{\textbf{Fig. S13.} Comparison of numerical and analytical phase velocities for the homogeneous acrylic plate. Phase velocity of the $A_0$ mode calculated using the numerical dispersion procedure (circles) and the analytical expression for flexural-wave phase velocity (solid line; Eq. 4).}
    \label{fig:supp-simValidation}
\end{figure*}
\vspace*{\fill}

\begin{figure*}[p]
    \centering
    \includegraphics[scale=0.9]{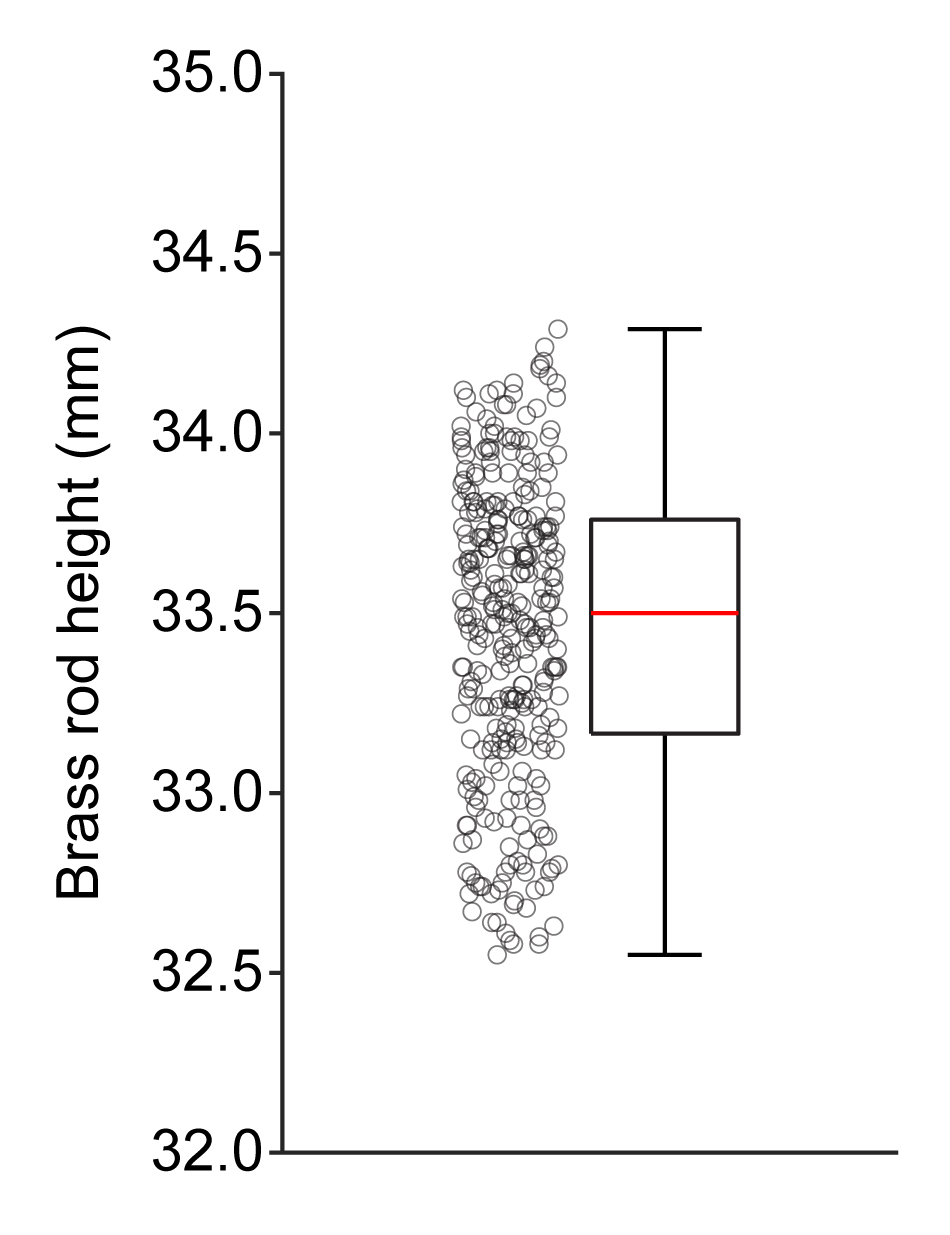}
    \caption{\textbf{Fig. S14.} Empirical measurements of brass rod heights ($n$ = 306) used to construct the metamaterial plate. Summary box plot of the distribution shown on the right (red: median, box: interquartile range, whiskers: maximum and minimum values).}
    \label{fig:ch6-supp-rodHeights}
\end{figure*}
\vspace*{\fill}

\begin{figure*}[p]
    \centering
    \includegraphics[scale=0.9]{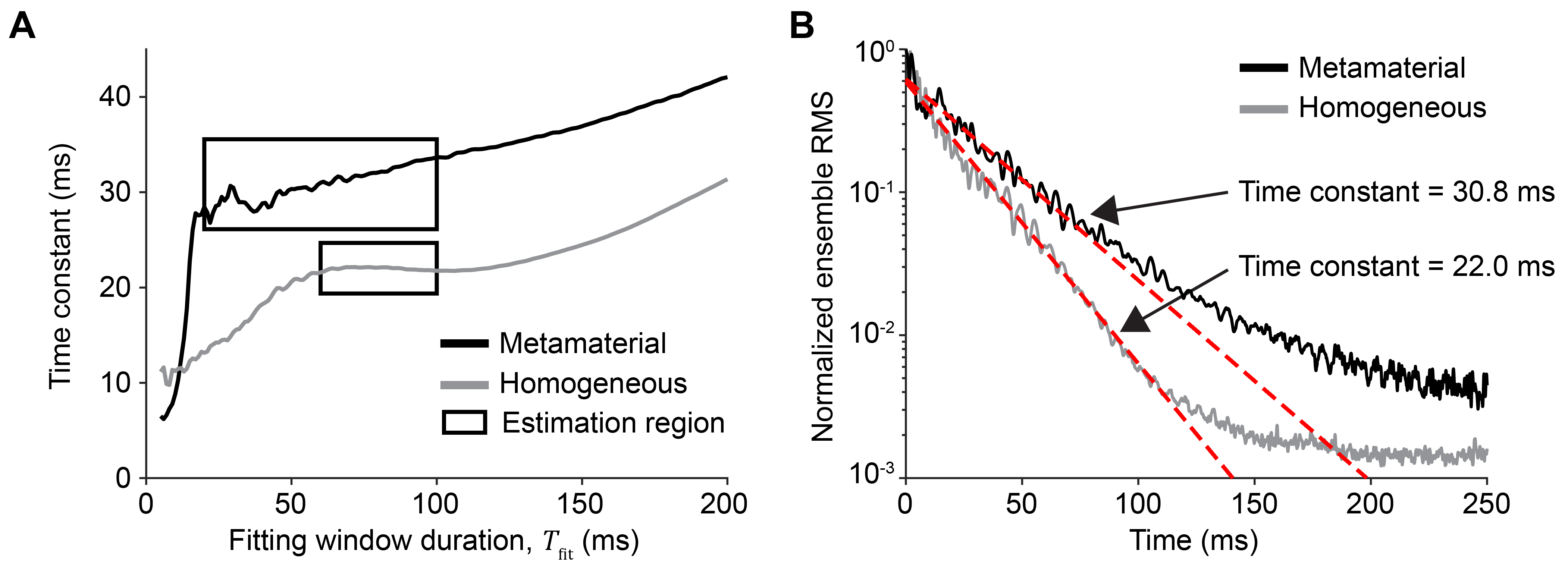}
    \caption{\textbf{Fig. S15. Attenuation time constants of the metamaterial and homogeneous acrylic plates.} (\textbf{A}) Estimated attenuation time constant $\tau$ as a function of fitting-window duration $T_\mathrm{fit}$ for the metamaterial (black) and homogeneous acrylic (gray) plates. Boxes indicate the ranges of fitting-window durations used to calculate the median attenuation time constant for each plate. (\textbf{B}) Normalized ensemble RMS responses computed from the measured Green's functions (solid lines) and corresponding exponential decay fits (red dashed lines). The estimated attenuation time constants were 30.8 ms for the metamaterial plate (black) and 22.0 ms for the homogeneous acrylic plate (gray).}
    \label{fig:supp-timeconstant}
\end{figure*}
\vspace*{\fill}

\end{document}